\documentclass[twocolumn]{aastex63}
\usepackage{CJK}
\usepackage{amsmath}
\newcommand{\sersic}{S\'{e}rsic}

\makeatletter

\newcommand{\Rmnum}[1]{\expandafter\@slowromancap\romannumeral #1@}
\makeatother
\defcitealias{1976ApJ+Faber2}{Faber--Jackson~(1976)}

\begin{document}
\begin{CJK*}{UTF8}{gkai}

  \title{The Carnegie-Irvine Galaxy Survey. \Rmnum{10}. Bulges in Stellar Mass-based Scaling Relations}

  \author[0000-0003-1015-5367]{Hua Gao (高桦)}
  \affiliation{Kavli Institute for Astronomy and Astrophysics, Peking University, Beijing 100871, China}
  \affiliation{Kavli Institute for the Physics and Mathematics of the Universe (WPI), The University of Tokyo Institutes for Advanced Study, The University of Tokyo, Kashiwa, Chiba 277-8583, Japan}
  \affiliation{Institute for Astronomy, University of Hawaii, 2680 Woodlawn Drive, Honolulu HI 96822, USA}

  \author[0000-0001-6947-5846]{Luis C. Ho}
  \affiliation{Kavli Institute for Astronomy and Astrophysics, Peking University, Beijing 100871, China}
  \affiliation{Department of Astronomy, School of Physics, Peking University, Beijing 100871, China}


  \author[0000-0001-5017-7021]{Zhao-Yu Li}
  \affiliation{Department of Astronomy, Shanghai Jiao Tong University, Shanghai 200240, China}

  \correspondingauthor{Hua Gao}
  \email{hgao.astro@gmail.com}

  \begin{abstract}
    We measure optical colors for the bulges of 312 disk galaxies from the Carnegie-Irvine Galaxy Survey and convert their previously available $R$-band structural parameters to stellar mass parameters. We also measure their average stellar mass surface density in the central 1\,kpc ($\Sigma_{1}$). Comparing the mass-based Kormendy relation with the original one based on flux, we find that the majority of the classifications into classical and pseudo bulges, as well as their overall statistical properties, remain essentially unchanged. While the bulge type classifications of the Kormendy relation are robust against stellar population effects, the mass-based classification criteria do produce better agreement between bulge structural properties and their stellar populations. Moreover, the mass-based Kormendy relation reveals a population of ultra-dense bulges akin to high-$z$ compact early-type galaxies, which are otherwise hidden in the original Kormendy relation. These bulges are probably relics of spheroids assembled in the early Universe, although for some we cannot rule out some contribution from secular growth. We confirm previous studies that $\Sigma_1$ correlates well with bulge surface densities.
  \end{abstract}


  \section{Introduction}
  \label{sec:introduction}

  First established in the 1980s \citep{1987ApJ+Djorgovski,1987ApJ+Dressler}, the fundamental plane is a tight relation between observables in size, luminosity, and kinematics of early-type galaxies that unifies its two earlier discovered projections, the \citetalias{1976ApJ+Faber2} relation between luminosity and stellar velocity dispersion and the \citet{1977ApJ+Kormendy2} relation between effective radius and surafce brightness. Besides its original usage as a distance indicator, it has provided much insight into the formation and evolution of early-type galaxies \citep{1987ApJ+Dressler, 1987nngp+Faber, 1988ASPC+Djorgovski, 1988ApJ+Lynden-Bell, 2013MNRAS+DOnofrio, 2017ApJ+DOnofrio}.  Rooted in the virial theorem, the fundamental plane and its projections have achieved notable success because they provide an empirical foundation to distinguish stellar systems of different physical nature. For instance, the fundamental plane has helped establish that spheroidal galaxies are physically distinct from giant ellipticals \citep{1985ApJ+Kormendy, 1987nngp+Kormendy, 1992ApJ+Bender, 1994ESOC+Binggeli}, and that a dichotomy exists between slow-rotator and fast-rotator ellipticals \citep{1997AJ+Faber}.  Recently, the Kormendy relation has played an increasingly important role as a diagnostic tool to distinguish classical bulges from pseudo bulges \citep{2009MNRAS+Gadotti,2017A&A+Neumann,2020ApJS+Gao}. Classical bulges share similar observational properties with ellipticals \citep{2004ARA&A+Kormendy,2008AJ+Fisher,2020ApJS+Gao} and therefore are often regarded as end products of violent processes such as galaxy mergers and coalescence of stellar clumps in high-$z$ disks \citep[e.g.,][]{1977egsp+Tmoore,2016ASSL+Bournaud,2019MNRAS+Tacchella}. Pseudo bulges, by contrast, are basically miniature disks masquerading as bulges in the center of a galaxy \citep{1993IAUS+Kormendy, 2004ARA&A+Kormendy, 2008AJ+Fisher, 2009ApJ+Fisher, 2020ApJS+Gao}. They originate from redistributed disk material by secular processes driven by non-axisymmetric mass distributions such as bars and spirals \citep{1981A&A+Combes, 1981seng.proc+Kormendy, 1982SAAS+Kormendy, 1990ApJ+Pfenniger, 1993RPPh+Sellwood, 2004ARA&A+Kormendy, 2014RvMP+Sellwood}.

  Our previous study showed that classical and pseudo bulges selected by the Kormendy relation have physical properties consistent with what we expect from their distinctive formation paths \citep{2020ApJS+Gao}. We performed multicomponent decomposition of $R$-band images of galaxies from the Carnegie-Irvine Galaxy Survey (CGS\@; \citealp{2011ApJS+Ho}) to measure accurate bulge parameters, and presented the statistics of classical and pseudo bulges based on classifications made using the Kormendy relation. Pseudo bulges generally are less concentrated and less prominent, and they have characteristically fainter surface brightness than classical bulges. Classical bulges exhibit an apparent ellipticity distribution skewed toward lower values compared with pseudo bulges, hinting that classical bulges are intrinsically rounder. In contrast with previous studies, we found that the distribution of the \citet{1968adga+Sersic} index of bulges is not bimodal, and that the two kinds of bulges cannot be well-separated by the widely adopted criterion of $n=2$ \citep[e.g.,][]{2008AJ+Fisher, 2016ASSL+Fisher}. We also found many fewer pseudo bulges in early-type disk galaxies, which helps alleviate the tension presented by the overbundance of massive, pure disk galaxies in a hierarchically clustered, $\Lambda$CDM Universe \citep[e.g.,][]{2008ASPC+Kormendy, 2010ApJ+Kormendy2,2010Natur+Peebles}.

  Despite the above success, we note that relying on single-band measurements has potential drawbacks. The influence of stellar population must be explored. Systematic variations in mass-to-light ratio ($M/L$) have been reported to cause the tilt of the fundamental plane with respect to the virial plane \citep[e.g.,][]{1987ApJ+Djorgovski, 1993ApJ+Renzini, 2001MNRAS+Kroupa, 2012Natur+Cappellari}. \citet{2013MNRAS+Cappellari1} found that replacing luminosity with dynamical mass brought the relation closer to the virial plane and reduced the scatter. This is also true for the case of stellar mass \citep{2009MNRAS+Hyde}. It is well known that pseudo bulges have younger stellar populations than classical bulges \citep[e.g.,][]{2009ApJ+Fisher,2016ASSL+Sanchez-Blazquez}, and some classical bulges may also have residual star formation \citep{2020MNRAS+Luo}. In light of these considerations, we expect that replacing surface brightness with stellar mass surface density in the Kormendy relation would provide a more fundamental, physical parameter space with which to investigate bulges.

  The structural properties of galaxies are closely related to their level of star formation activity. Early-type galaxies, being centrally concentrated, are also predominantly quiescent in the local Universe. Various structural parameters have been explored as discriminators of star-forming and passive galaxies, including \sersic{} index, surface stellar mass density, and stellar mass divided by the half-light radius \citep[e.g.,][]{2003MNRAS+Kauffmann, 2006MNRAS+Kauffmann, 2008ApJ+Bell, 2008ApJ+Franx, 2011ApJ+Wuyts, 2012ApJ+Bell}. Inspired by these studies, \cite{2012ApJ+Cheung} first introduced the parameter $\Sigma_{1}$, the average stellar mass surface density in the central 1\,kpc, to trace the star formation status of galaxies, comparing its efficacy relative to other structural parameters. Subsequently, $\Sigma_{1}$ enjoyed wide application, especially in the context of scaling relations to study galaxy quenching and structural transformation \citep{2013ApJ+Fang,2019MNRAS+Woo,2017MNRAS+Woo}. By design, this parameter closely traces bulge properties, as typical bulge size is $\sim1\,\mathrm{kpc}$ \citep{2008AJ+Fisher, 2020ApJS+Gao}. The increase of $\Sigma_{1}$ as galaxies evolve from the blue cloud to the red sequence is often associated with compaction processes such as violent disk instabilities and mergers \citep{2015MNRAS+Zolotov,2016MNRAS+Tacchella}, during which bulges also grow. Therefore, it is quite natural to extend its application to study bulges \citep{2020MNRAS+Luo,2020ApJ+Yesuf}. The classifications of bulge type based on $\Sigma_{1}$ agree with those based on the Kormendy relation. Most importantly, $\Sigma_{1}$ is simple and economical to measure compared with detailed bulge structural parameters that require adequate spatial resolution and time-consuming, often highly uncertain bulge-to-disk decompositions. Nevertheless, some information would be lost from averaging the central mass distribution, and contamination from other structural components is inevitable.

  The success of working with stellar mass in studying galaxy evolution motivates us to extend the measurements of the $R$-band bulge structural parameters of 320 CGS disk galaxies \citep{2019ApJS+Gao}. This contribution focuses on measuring the stellar mass parameters of these objects based on their optical colors. We then use these measurements to construct a revised mass-based Kormendy relation of bulges and ellipticals and compare classifications of bulge types with those obtained with the original Kormendy relation \citep{2020ApJS+Gao}. We will also evaluate the feasibility of using $\Sigma_{1}$ as an additional diagnostic to study bulge and galaxy evolution.

  Bulge colors can be measured through multiband bulge-to-disk decompositions, as conducted, for instance, in the MegaMorph project \citep[e.g.,][]{2011ASPC+Bamford,2014MNRAS+Vika,2016MNRAS+Kennedy}. Their approach can increase the signal-to-noise ratio of the data and self-consistently account for the flux contribution from other components. Due to the large sample size, their surface brightness models are relatively simple for practical reasons. The multicomponent models of the CGS sample are much more detailed and complicated than those adopted in large data sets \citep{2018ApJ+Gao,2019ApJS+Gao}, rendering it extremely computationally expensive and unfeasible to perform multiband fitting for our study. Moreover, MegaMorph's assumption that the structural parameters vary as polynomial functions of wavelength lacks physical basis. Instead, we adopt the pragmatic strategy of estimating the bulge color using aperture photometry in different bands, defining the aperture size to include only the region where the bulge surface brightness dominates.  Although we do not remove the disk rigorously, its contamination can be minimized by proper choice of aperture size.

  This study uses the same sample as that presented in \citet{2019ApJS+Gao}. Sample selection and data quality are described in Section~\ref{sec:sample-data}. Section~\ref{sec:methodology} discusses the methodology to measure bulge color and stellar mass. Results and discussions are given in Sections~\ref{sec:result}~and~\ref{sec:discussion}. We summarize the paper in Section~\ref{sec:summary}.

  \section{Sample and Data}
  \label{sec:sample-data}

  Our sample, introduced in \citet{2019ApJS+Gao,2020ApJS+Gao}, is a subset of CGS (\citealt{2011ApJS+Ho}), which comprises bright ($B_{T}\leq 12.9\,\mathrm{mag}$) galaxies in the southern hemisphere without any selection on morphology, size, or environment. High-quality images in $B$, $V$, $R$, and $I$, along with isophotal parameters, were presented in \cite{2011ApJS+Li}. The images have a field-of-view of $8\farcm9\times8\farcm9$, a median seeing of $\sim 1\arcsec$, and a median surface brightness sensitivity of 26.9\,mag~arcsec$^{-2}$ in $B$ and 26.4\,mag~arcsec$^{-2}$ in $R$. The median luminosity distance of the sample is 26\,Mpc, almost all galaxies are within 80\,Mpc \citep{2019ApJS+Gao}, and the median seeing ($1\arcsec$) of R-band images corresponds to $\sim 0.1\,\mathrm{kpc}$ and $\sim 0.4\,\mathrm{kpc}$ at these distances, respectively. Note that the typical bulge half-light radius is $\sim$1\,kpc, which is well resolved in the CGS images. This study makes use only of the $B$ and $R$ image products.

  In previous papers of this series, we performed two-dimensional bulge-to-disk decompositions of the disk (lenticular and spiral) galaxies of the CGS sample in the $R$ band \citep{2018ApJ+Gao, 2019ApJS+Gao} and studied the statistical properties of classical and pseudo bulges using the measurements \citep{2020ApJS+Gao}. The bulge-to-disk decompositions were performed on a subset of CGS galaxies selected by morphological type index $-3\leq T\leq 9.5$ and inclination angle $i\leq 70\arcdeg$. We complemented the sample with a handful of misclassified ellipticals and removed objects whose decomposition was unsuccessful, arriving at a final sample of 320 galaxies. The decomposition strategy, detailed in Section~3 of \citet{2019ApJS+Gao}, considered nuclei, nuclear/inner lenses, inner rings, bars, and disk breaks. Our modeling strategy treated nuclear rings and nuclear bars as part of the bulge component, while other features such as spiral arms, outer lenses, and outer rings were omitted from the fits because they are not crucial for accurate bulge measurements according to the experiments in \citet{2017ApJ+Gao}. Active galaxy nuclei and nuclear star clusters are modeled as point sources, in order to remove their contamination to other galaxy components. The error budget of the bulge parameters includes the uncertainties from sky measurements and model assumptions.

  This study builds on the 320 disk galaxies with accurate bulge structural parameters. We succeed in deriving optical $B-R$ colors for the bulges of all but eight of the galaxies. We measure $\Sigma_{1}$ for the 312 disk galaxies, as well as an additional 83 CGS ellipticals serving as a reference sample. The structural parameters of the ellipticals were already presented in \citet{2020ApJS+Gao}.

  \section{Methodology}
  \label{sec:methodology}


  Based on the two-dimensional best-fit models of the galaxies from \citet{2019ApJS+Gao}, to estimate bulge colors we define a ``bulge aperture'' as the set of image pixels where the flux of the bulge dominates over that of other components. In order to secure a reasonably large aperture to estimate bulge colors for most of the galaxies, we select pixels having a bulge flux threshold at least 25\% brighter than all the other components combined. Next, we remove pixels that are affected by dust lanes and foreground sources. \citet{2019ApJS+Gao} already paid careful attention to mask manually central dust lanes when performing bulge-to-disk decompositions. However, we do not expect that the dust lanes were perfectly removed. Therefore, we take extra precaution to further minimize dust extinction effects by performing $1.5\sigma$ clipping of pixels that show irregular colors compared with other pixels at the same galactic radius. We further remove pixels whose flux contains non-zero contribution from active galactic nuclei or nuclear star clusters, if present, in order to minimize their impact on estimating bulge colors. Finally, we remove pixels inside 3 times the seeing disk; for galaxies with small bulges, we remove pixels inside one seeing disk to preserve sufficient pixels to estimate their colors. Such galaxies are mostly late-type spirals with pseudo bulges, whose centers are often plagued by severe dust extinction, which further exacerbates the situation. Eventually, we omit eight galaxies from the sample of \citet{2019ApJS+Gao} because there are not enough pixels left to ensure robust color measurement.

  We calculate an integrated $B-R$ color over the above-selected pixels in the $B$ and $R$ images and treat it as the bulge color. The $R$-band stellar mass-to-light ratio follows from
  \begin{equation}
    \label{eq:ML}
    \log \left(M/L\right)_{R}=0.683(B-R)-0.523,
  \end{equation}
  with a typical error of $\sim 0.1\,\mathrm{dex}$ \citep{2003ApJS+Bell}. We then derive stellar mass and average stellar mass surface density within the effective radius $r_e$ ($\langle\Sigma_{e}\rangle$) of the bulges. We perform the same calculations for the ellipticals, which already have available colors \citep{2011ApJS+Li} and structural parameters \citep{2020ApJS+Gao}. The uncertainties of $B-R$ include photometric errors in both bands, uncertainties in defining bulge aperture introduced by the bulge structural parameters, and dispersion of colors in the bulge aperture caused by possible radial gradients. The latter may reflect intrinsic color gradients and/or contamination from other structural components (e.g., disk). It is reassuring that for most cases the photometric errors dominate the error budget. We propagate the uncertainties of $B-R$, bulge magnitudes, and  Equation~(\ref{eq:ML}) to the uncertainties of the stellar mass parameters. The mean uncertainties of bulge stellar masses and average effective stellar mass surface density are 0.19\,dex and 0.18\,dex, respectively (Table~\ref{tab:measurements}).

  \begin{deluxetable*}{lRRRR}
    \tablecaption{Colors and Stellar Mass Parameters\label{tab:measurements}}
    \tablehead{\colhead{Name} & \colhead{$B-R$} & \colhead{$\log M_{\star}$} & \colhead{$\log \langle\Sigma_e\rangle$} & \colhead{$\log \Sigma_1$} \\
      \colhead{} & \colhead{(mag)} & \colhead{($M_{\sun}$)} & \colhead{($M_{\sun}\,\mathrm{kpc}^{-2}$)} & \colhead{($M_{\sun}\,\mathrm{kpc}^{-2}$)} \\
      \colhead{(1)} & \colhead{(2)} & \colhead{(3)} & \colhead{(4)} & \colhead{(5)}}

    \startdata
    ESO 027--G001 & 0.89\pm 0.23 &  8.66\pm 0.21 &  9.60\pm 0.20 &  8.87\pm 0.19\\
    ESO 121--G026 & 1.47\pm 0.10 &  9.92\pm 0.14 &  9.82\pm 0.13 &  9.63\pm 0.12\\
    ESO 137--G034 & 1.39\pm 0.03 & 10.40\pm 0.12 &  9.68\pm 0.11 &  9.77\pm 0.10\\
    ESO 138--G010 & 1.31\pm 0.24 &  9.29\pm 0.30 &  8.78\pm 0.27 &  8.94\pm 0.19\\
    ESO 185--G054 & 1.44\pm 0.02 & 11.69\pm 0.12 &  8.55\pm 0.13 & 10.13\pm 0.10\\
    ESO 186--G062 & 1.20\pm 0.26 &  9.51\pm 0.26 &  8.19\pm 0.27 &  8.60\pm 0.19\\
    ESO 213--G011 & 1.38\pm 0.09 &  9.74\pm 0.16 &  8.96\pm 0.15 &  9.28\pm 0.12\\
    ESO 221--G026 & 1.05\pm 0.23 &  9.93\pm 0.27 &  9.64\pm 0.25 &  9.44\pm 0.19\\
    ESO 221--G032 & 0.97\pm 0.06 &  9.09\pm 0.16 &  9.82\pm 0.14 &  9.06\pm 0.11\\
    ESO 269--G057 & 1.48\pm 0.23 & 10.58\pm 0.21 & 10.09\pm 0.20 & 10.03\pm 0.19\\
    ESO 271--G010 & 0.87\pm 0.06 &  8.37\pm 0.13 &  8.31\pm 0.12 &  8.60\pm 0.11
    \enddata
    \tablecomments{Col.~(1): Galaxy name. Col.~(2): Color of bulge or elliptical; data for ellipticals are from \citet{2011ApJS+Li}. Col.~(3): Stellar mass of bulge or elliptical. Col.~(4): Mean stellar mass surface density within the effective radius of bulge or elliptical. Col.~(5): Mean stellar mass surface density within a radius of 1 kpc. \\
      (Table~\ref{tab:measurements} is published in its entirety in machine-readable format. A portion is shown here for guidance regarding its form and content.)}
  \end{deluxetable*}

  Measurements of $\Sigma_{1}$ of disk and elliptical galaxies are straightforward. Given the isophotal analysis provided by \citet{2011ApJS+Li}, we integrate the $B$-band and $R$-band surface brightness profiles from the galaxy center outward to calculate the $B-R$ color and $\Sigma_{1}$, using Equation~(\ref{eq:ML}) to estimate $M/L$. As with the measurements of stellar mass parameters, we propagate the uncertainties from the photometric errors, interpolation errors in surface brightness profiles, and errors in Equation~(\ref{eq:ML}) into the uncertainties of $\Sigma_{1}$. The mean uncertainty of $\Sigma_1$ is 0.15\,dex (Table~\ref{tab:measurements}). Note that we exclude the hosts of known active galactic nuclei from our sample when measuring $\Sigma_{1}$.

  \begin{figure*}
    \epsscale{1.15}
    \plotone{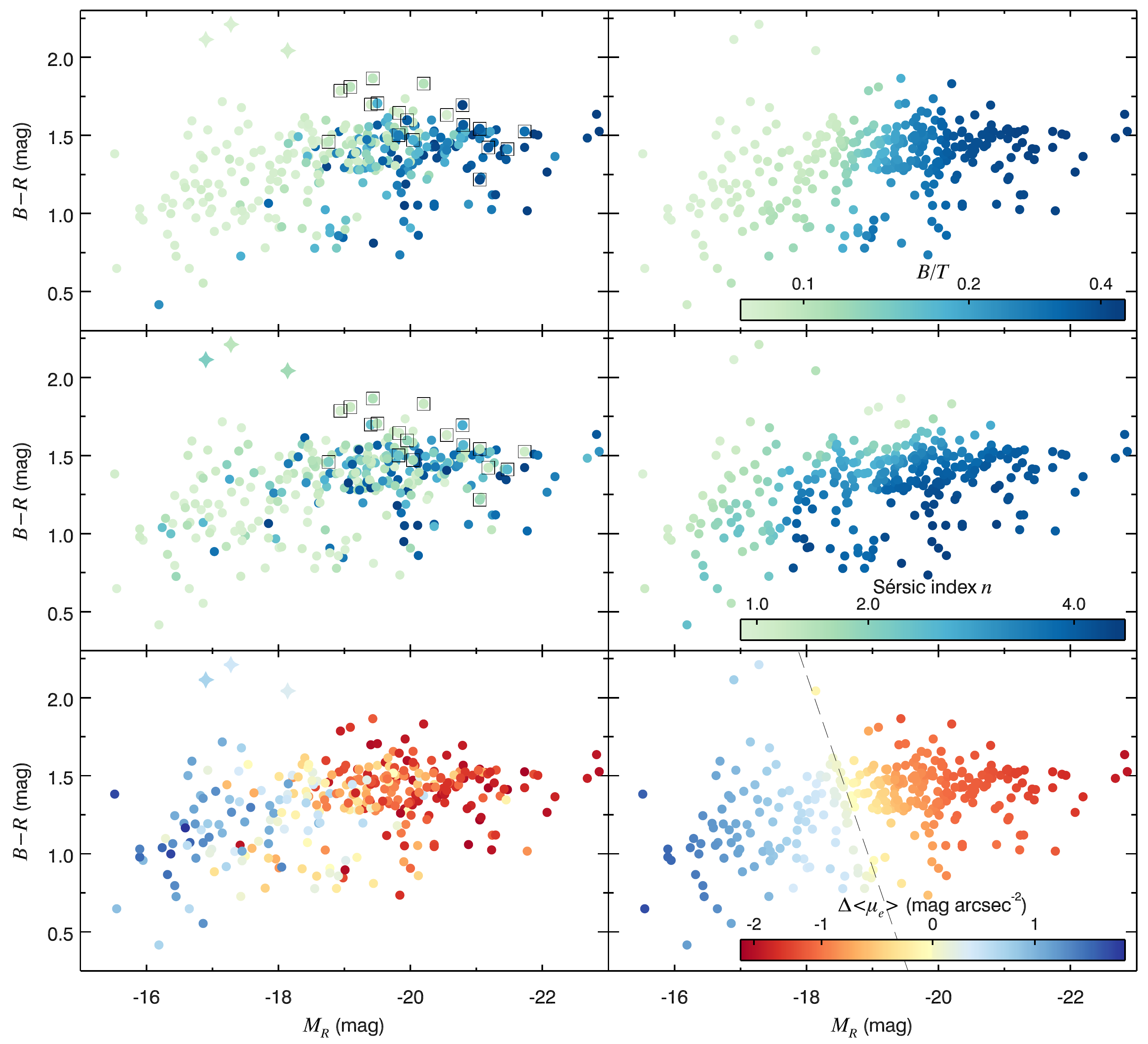}
    \caption{Color--magnitude diagram of bulges. From top to bottom, the symbols are color-coded according to the bulge-to-total ratio $B/T$, \sersic{} index $n$, and residual average effective surface brightness $\Delta \langle \mu_e \rangle$. The left panels show the original data, and the right panels are the LOESS-smoothed version. The symbols surrounded by large open squares are ultra-dense bulges (Section~\ref{sec:ultra-dense-bulges}). The stars in the left panels mark three outliers (NGC~4947, NGC~5188, and NGC~7689), which are discussed in the main text. The black dashed line in the bottom panel is the demarcation between classical and pseudo bulges classified using the Kormendy relation, obtained, for illustration purposes only, by fitting the data with $\Delta \langle \mu_e \rangle \approx 0$ (yellowish symbols). \label{fig:col_mag}}
  \end{figure*}

  \begin{figure*}
    \epsscale{1.17}
    \plotone{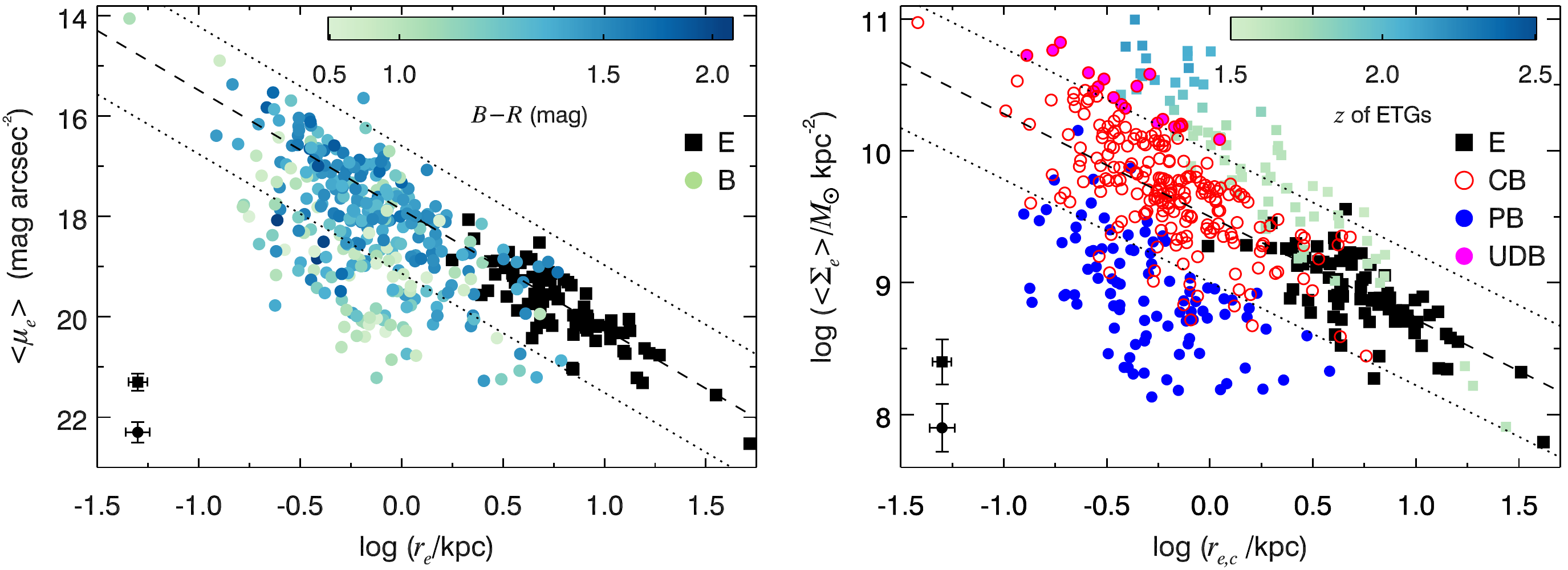}
    \caption{The Kormendy relation of ellipticals and bulges (labeled E and B, respectively), shown in the traditional flux-based units in the $R$ band (left) and stellar mass units (right). The dashed and dotted lines represent the best-fit relation and its $3\sigma$ boundary of ellipticals (filled black squares). The filled circles in the left panel are bulges color-coded based on their $B-R$ color. In the right panel, the filled blue circles and open red circles are pseudo bulges (labeled PB) and classical bulges (labeled CB), respectively, classified using the flux-based Kormendy relation; the filled magenta circles are ultra-dense bulges (labeled UDB\@; Section~\ref{sec:ultra-dense-bulges}); the smaller greenish/blueish squares are high-$z$ ETGs color-coded according to their redshift and LOESS-smoothed. Representative error bars are given in the lower-left corner. \label{fig:KR}}
  \end{figure*}

  \section{Results}
  \label{sec:result}

  \subsection{Strategies for Classifying Bulges}
  \label{sec:urge-replace-surface}

  Figure~\ref{fig:col_mag} presents the color--magnitude diagram of bulges, color-coded, respectively, using the bulge-to-total flux ratio ($B/T$), \sersic{} index ($n$), and residual average effective surface brightness relative to the lower $3\sigma$ boundary of the Kormendy relation ($\Delta \langle \mu_{e} \rangle$; Figure~\ref{fig:KR}). The structural parameters are taken from \citet{2020ApJS+Gao} and were measured in the $R$ band. To minimize the effects of object overlay and aid visual inspection of the average trends, we smoothed the data by locally weighted regression (LOESS\@; \citealp{1988J.AM.STAT.ASSOC+Cleveland}).\footnote{We adopt the implementation of the two-dimensional LOESS method by \citet{2013MNRAS+Cappellari2}, who kindly provide the code in \url{http://purl.org/cappellari/idl}.} Three extremely red ($B-R>2$\,mag) but small bulges (NGC~4947, NGC~5188, and NGC~7689) are readily recognized as outliers to the majority (stars in Figure~\ref{fig:col_mag}). We are confident that NGC~5188 has an unreliable color, because its center is complicated by intricate dust features that are beyond the capacity of manual masking and pixel clipping. NGC~4947 has well-defined dust lanes that were masked during measuring its bulge color, but we cannot exclude the possibility that its exceptionally red color is caused by continuous dust extinction. Nothing particularly suspicious stands out about the bulge color of NGC~7689. We do not remove these three sources from the analysis, as they do not affect the interpretation.

  Comparison of the three color--magnitude diagrams shows that the Kormendy relation is superior in classifying bulges compared with single structural parameters such as $B/T$ or $n$, because only $\Delta\mu_{e}$ presents a systematic trend from red, luminous systems, characteristic of typical classical bulges, to blue, faint ones, as expected of pseudo bulges. Notably, $n$ gives the worst performance, showing no discernible trends with either bulge color or luminosity. The bulge-to-total ratio does a better job, although the levels of constant $B/T$ do not arrange bulges into a sequence that resembles the transition we expect from classical to pseudo bulges. These results echo the findings of \citet{2020ApJS+Gao}, who conclude that $n$ is not an effective indicator of bulge properties, whereas the Kormendy relation is more robust. Pseudo bulges selected according to the Kormendy relation exhibit a wider range of colors than classical bulges. Indeed, Figure~\ref{fig:col_mag} reveals a population of pseudo bulges with colors as red as classical bulges ($B-R \approx 1.5$ mag). This suggests that incorporating stellar population information into the Kormendy relation may be able to improve the classification further. Figure~\ref{fig:KR} illustrates how bulge colors are distributed in the Kormendy relation (left panel). We expect that some of the red pseudo bulges (having large $M/L$) close to the lower $3\sigma$ boundary will be classified as classical bulges once we replace surface brightness with mass surface density, and vice versa.

  \subsection{The Mass-based Kormendy Relation}
  \label{sec:mass_KR}

  Figure~\ref{fig:KR} directly contrasts the traditional $R$-band Kormendy relation \citep{2020ApJS+Gao} with our new stellar mass-based version\footnote{Unlike \cite{2020ApJS+Gao}, the effective radius here has been circularized by $r_{e,c} = r_e \sqrt{1-\epsilon}$, with $\epsilon$ the apparent ellipticity of the best-fit bulge model.}, which replaces the average effective surface brightness ($\langle\mu_e\rangle$) with the average effective stellar mass surface density ($\log\langle\Sigma_{e}\rangle$). In addition to CGS ellipticals and bulges, we also overlay for comparison the sample of massive, high-redshift ($z \approx 1.5\textrm{--}2.5$) early-type galaxies (ETGs), whose structural parameters were compiled by \citet{2013ApJ+Huang2} from the literature \citep{2007ApJS+Scarlata, 2007ApJS+Scoville, 2009A&A+Tasca, 2011MNRAS+Conselice, 2011ApJ+Damjanov, 2011ApJS+Grogin, 2012ApJ+Papovich, 2012ApJ+Szomoru}. As in \citet{2020ApJS+Gao}, we use the best-fit relation of the ellipticals to separate classical and pseudo bulges:
  \begin{eqnarray}
    \label{eq:kr-ell}
    \log (\langle \Sigma_{e} \rangle / M_{\sun}\,\mathrm{kpc}^{-2})&=&\left(-0.78\pm0.07\right)\log (r_{e,c}/\mathrm{kpc}) \nonumber \\
    & & +\left(9.50\pm 0.05\right),
  \end{eqnarray}
  which has a scatter of 0.17\,dex in $\log\langle\Sigma_{e}\rangle$. We define pseudo bulges as those with $\log\langle\Sigma_{e}\rangle$ smaller than the lower $3\sigma$ boundary of the above relation,
  \begin{equation}
    \label{eq:pb-def}
    \log(\langle\Sigma_{e}\rangle/M_{\sun}\,\mathrm{kpc}^{-2})<-0.78\log(r_{e,c}/\mathrm{kpc})+8.99.
  \end{equation}
  The rest are classical bulges.

  As expected, in Figure~\ref{fig:KR} we notice bulges classified as one type using the original Kormendy relation cross over to the other category once we account for the effects of $M/L$. Among the original sample of 94 pseudo bulges, 15 are reclassified as classical bulges according to the mass-based Kormendy relation; as for classical bulges, we relabel 12 of 218 as pseudo bulges. The consistency rate is $91\%$. Thus, as discussed in the next section, the overall statistics of the two bulge populations are not significantly changed.

  \subsubsection{Better Agreement between Structural and Stellar Population Properties of Bulges}
  \label{sec:str-ste-pop}

\begin{figure*}
    \epsscale{1.15}
    \plotone{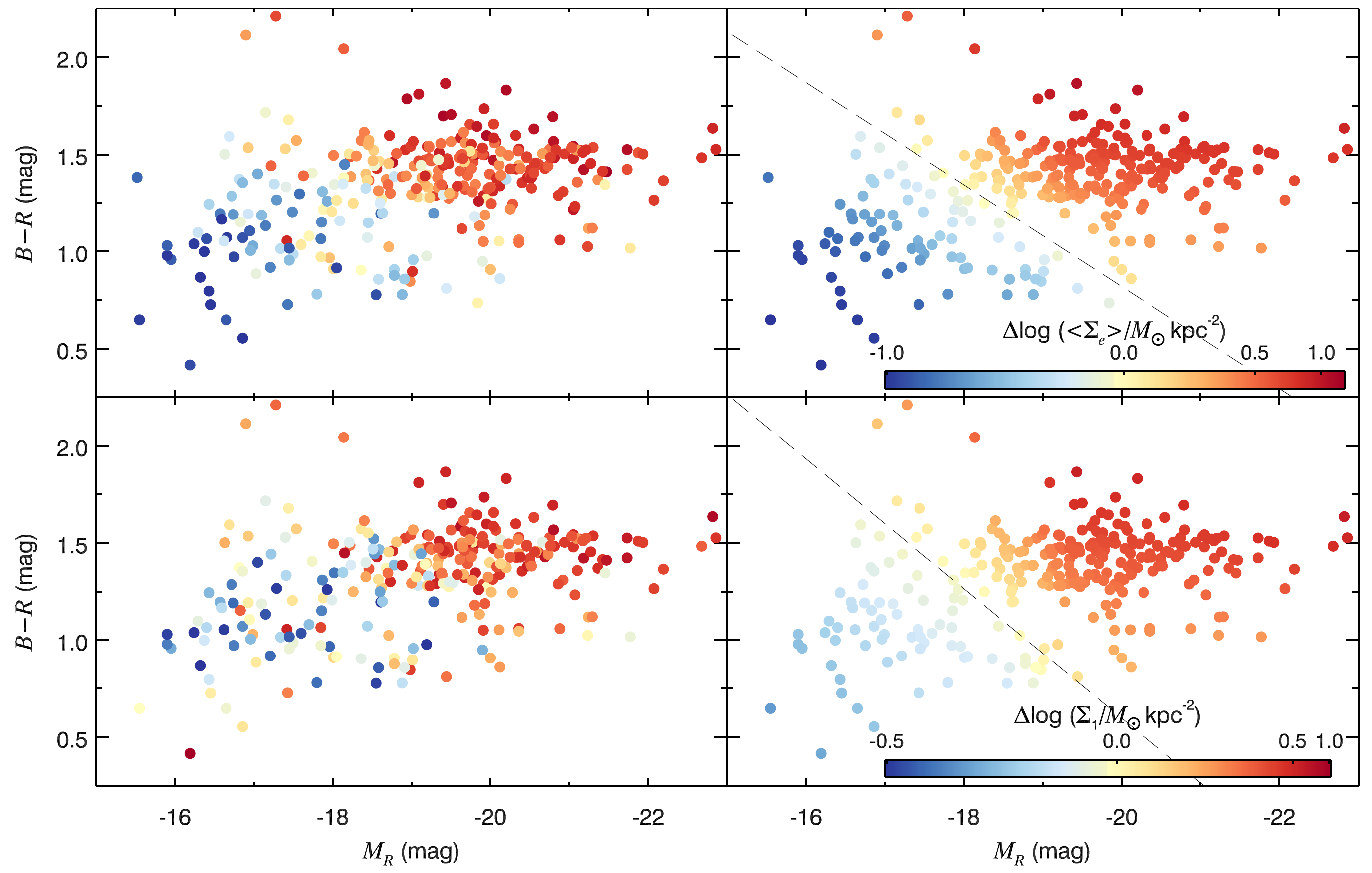}
    \caption{Color--magnitude diagram of bulges. The symbols are color-coded according to (top) $\Delta \log \langle \Sigma_e \rangle$ and (bottom) $\Delta \log \Sigma_1$. The left panels show the original data, and the right panels are the LOESS-smoothed version. The black dashed lines in the right panels are the demarcation between classical and pseudo bulges classified using (top) the mass-based Kormendy relation and (bottom) the $M_{\star}$--$\Sigma_1$ relation (Section~\ref{sec:sigm-class-bulg}), obtained, for illustration purposes only, by fitting the data with $\Delta \log \langle \Sigma_e \rangle \approx 0$ and $\Delta \log \Sigma_1 \approx 0$, respectively (yellowish symbols). \label{fig:col_mag_mass}}
  \end{figure*}

  Bulge classifications based on structural and stellar population properties  often yield inconsistent results. This is already seen in Figure~\ref{fig:col_mag} and described in Section~\ref{sec:urge-replace-surface}. Some pseudo bulges appear to have colors comparable to those of typical classical bulges. Figure~\ref{fig:col_mag_mass} gives another view of the bulge color--magnitude diagram, now with the bulges color-coded by their distance to the lower $3\sigma$ boundary of the mass-based Kormendy relation ($\Delta\langle\Sigma_{e}\rangle$). Comparison with Figure~\ref{fig:col_mag} shows that the average demarcation line of classical and pseudo bulges tilts counter-clockwise when the classifications are made using the mass-based relation (Figure~\ref{fig:col_mag_mass}). This allows more red bulges to enter the domain of classical bulges, thereby alleviating the tension between bulge structural and stellar population properties.

  \subsubsection{Updated Statistical Properties of Bulges}
  \label{sec:stat-prop-class}

  As mentioned in Section~\ref{sec:introduction}, based on the classifications made using the $R$-band Kormendy relation, pseudo bulges generally have less concentrated (lower $n$) light profiles, are less prominent (smaller $B/T$), and have fainter average surface brightness (larger $\langle\mu_e\rangle$) than classical bulges. Classical bulges show an apparent ellipticity distribution skewed toward lower values compared to pseudo bulges, since they are intrinsically rounder objects. We also found that the \sersic{} index distribution is not bimodal, and that the two kinds of bulges cannot be well separated at $n \approx 2$. Bulges classified using the Kormendy relation naturally show well-separated $\langle\mu_e\rangle$ distributions for classical and pseudo bulges. Our decomposition generally yielded less conspicuous bulges (smaller $B/T$) than previously measured, for both classical and pseudo bulges, on account of our more detailed and more accurate treatment of galactic subcomponents \citep{2017ApJ+Gao}.

\begin{figure*}
    \epsscale{1.15}
    \plotone{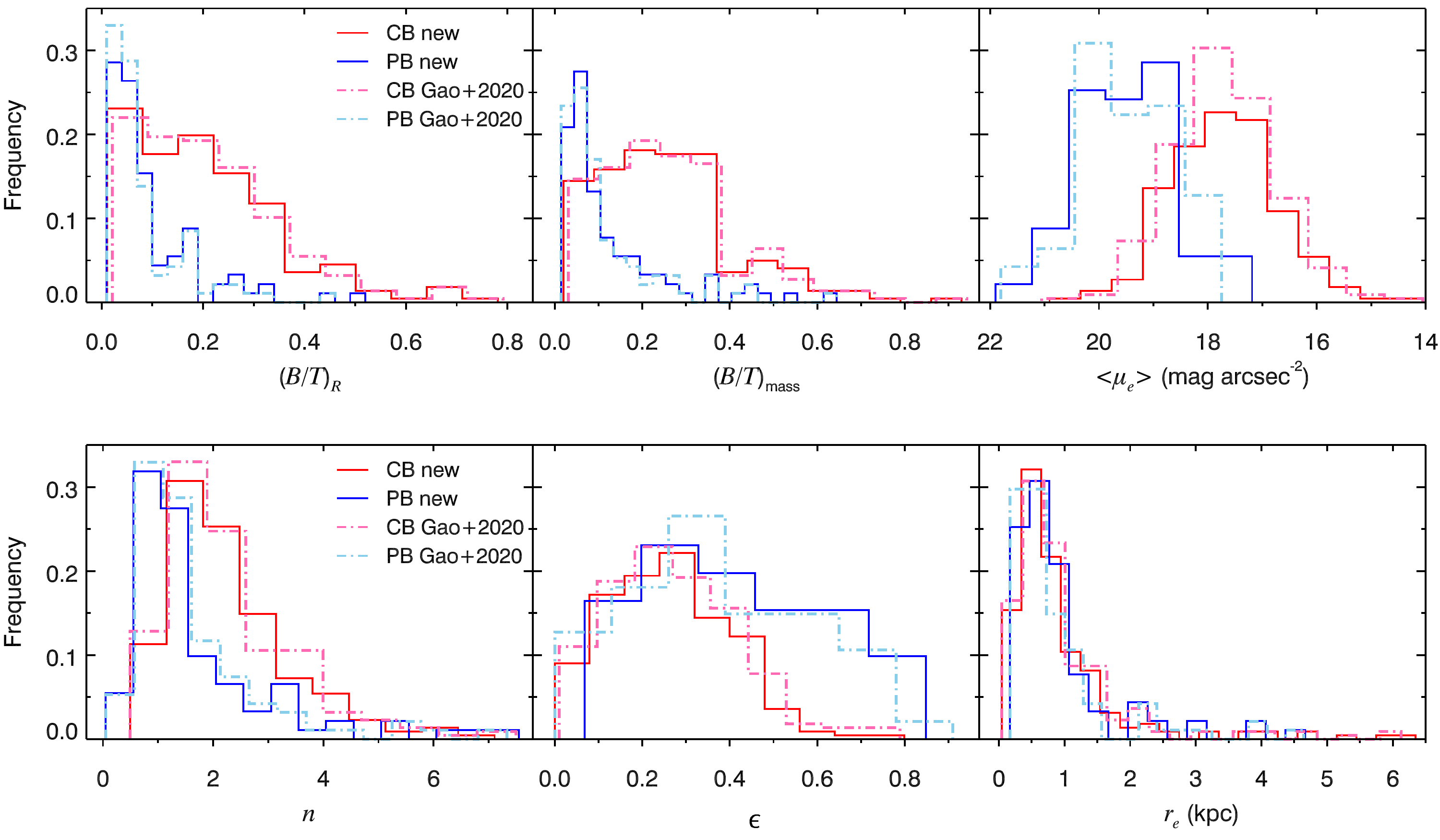}
    \caption{Comparison of structural parameters of classical (reddish) and pseudo (blueish) bulges classified using the original (dot-dashed histograms; \citealt{2020ApJS+Gao}) and the new (solid histograms) mass-based Kormendy relation.  From left to right, top to bottom, the panels display distributions of $R$-band bulge-to-total flux ratio, bulge-to-total mass ratio, average effective surface brightness ($\langle\mu_e\rangle$), \sersic{} index ($n$), apparent ellipticity ($\epsilon$), and effective radius ($r_e$) of the bulge. \label{fig:hist_comp}}
  \end{figure*}

  In light of the new classification criterion presented in this study, Figure~\ref{fig:hist_comp} reassesses the overall statistical properties of the structural parameters of classical and pseudo bulges classified using the original Kormendy relation (dot-dashed histograms) and the new mass-based version (solid histograms). The revision in classification criteria does not significantly alter the overall statistics of the two populations, apart from the tendency for the bulge-to-total mass ratios of the two types to be better segregated than the bulge-to-total flux ratios, now forming a nearly bimodal distribution.  And the two types now perhaps show a somewhat more noticeable separation in terms of \sersic{} index.  Overall, however, the conclusions in \citet{2020ApJS+Gao} largely still hold. This reinforces our previous conclusion that the Kormendy relation is already a robust tool to classify the two populations of bulges. Switching to stellar mass hardly affects the overall statistics.

  \begin{figure}
    \epsscale{1.18}
    \plotone{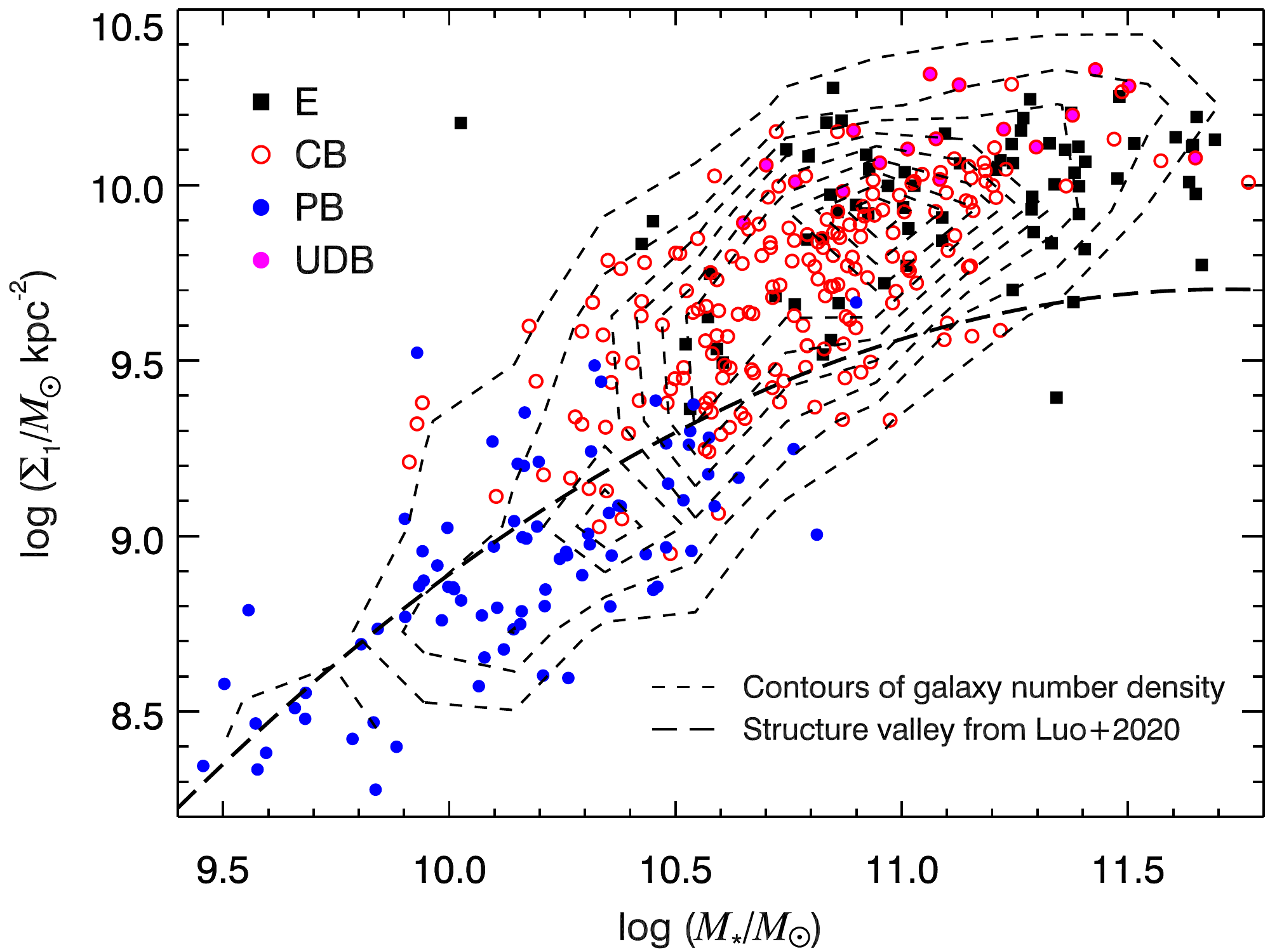}
    \caption{Correlation between $\Sigma_{1}$ and $M_{\star}$ of CGS ellipticals and disk galaxies. Black filled squares are ellipticals; filled circles are disk galaxies.  The hosts of classical bulges are in red, and the hosts of pseudo bulges, as selected by Equation~(\ref{eq:pb-def}), are in blue; the hosts of ultra-dense classical bulges defined by Equation~(\ref{eq:udb-def}) are in magenta.  The contours show the distribution of all galaxies. The red sequence is readily recognizable but the blue cloud is sparsely sampled.  The black dashed line is the structural valley from \citet{2020MNRAS+Luo}.\label{fig:mass_sigma1}}
  \end{figure}

  \begin{figure*}
    \epsscale{1.15}
    \plotone{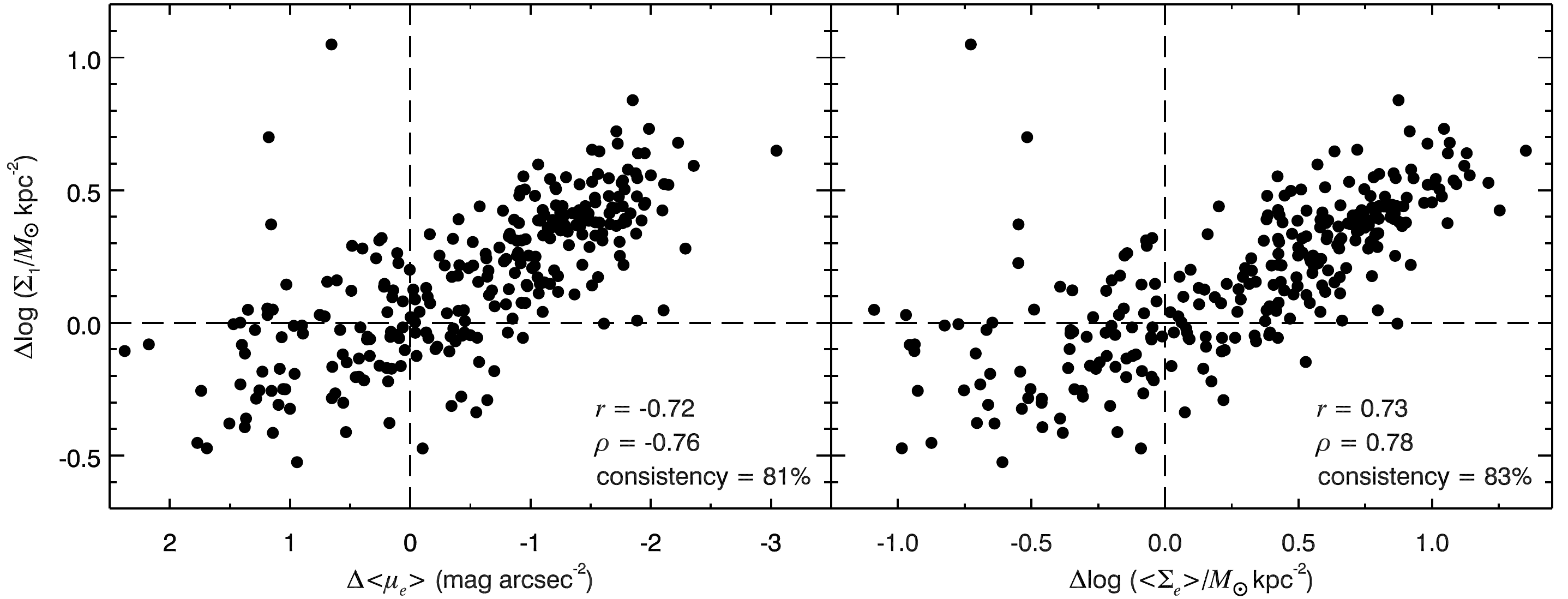}
    \caption{Correlation between $\Delta\log\Sigma_{1}$ and bulge $\Delta\langle\mu_{e}\rangle$ (left) and bulge $\Delta\log \langle\Sigma_{e}\rangle$ (right) for disk galaxies. The Pearson correlation coefficient ($r$) and Spearman's rank correlation coefficient ($\rho$) are given in the lower-right corner. Consistency is defined as the fraction of bulges in the first and third quadrants.
      \label{fig:delta_sigma1_mu}}
  \end{figure*}

  \subsubsection{Ultra-dense Bulges}
  \label{sec:ultra-dense-bulges}

  We draw attention to a population of classical bulges located above the upper $3\sigma$ boundary of the best-fit relation of the CGS ellipticals (magenta filled circles; Figure~\ref{fig:KR}b). The threshold
  \begin{equation}
    \label{eq:udb-def}
    \log(\langle\Sigma_{e}\rangle/M_{\sun}\,\mathrm{kpc}^{-2})>-0.78\log(r_{e,c}/\mathrm{kpc})+10.01
  \end{equation}
  selects 19 such ultra-dense bulges. This population is more prominent in the mass-based relation than in the original Kormendy relation, where only one such case stands out (Figure~\ref{fig:KR}a). Ultra-dense bulges have mass surface densities higher than those of $z\approx 0$ ellipticals but are comparable to those of $z \ga 1.5$ ETGs. As shown in Figure~\ref{fig:col_mag}, these objects are not necessarily the most prominent (large $B/T$ or bulge luminosity) or the most concentrated (large $n$) bulges, but they are the reddest at a given bulge luminosity. As discussed in Section~\ref{sec:origin-ultra-dense}, ultra-dense bulges may be well-preserved high-$z$ red nuggets or bulges with an early assembly history \citep{2015ApJ+Graham,2018ApJ+Gao}.

  \subsection{Correlation between $\Sigma_1$ and Bulge Parameters}
  \label{sec:sigm-class-bulg}

  \citet{2020MNRAS+Luo} investigated the feasibility of using $\Sigma_{1}$ to classify bulges, and concluded that this parameter can produce bulge classifications consistent with those derived from the Kormendy relation. They also stressed the importance of removing the mass dependence in $\Sigma_{1}$ when separating the two bulge populations. Our ability to revisit these issues is somewhat hampered by the under-representation of low-mass galaxies in CGS, as well as the overall small-number statistics of the survey, which preclude us from robustly defining the structural valley and measuring $\Delta\Sigma_{1}$ for the sample in a self-consistent manner, as shown in Figure~\ref{fig:mass_sigma1}\footnote{The figure omits a few outliers for better visualization.}. For the sake of convenience, we adopt the formulation of the structural valley from \citet{2020MNRAS+Luo} and calculate $\Delta\log \Sigma_{1}=\log \Sigma_{1}+0.275(\log M_{\star})^{2}-6.445\log M_{\star}+28.059$.


  Figure~\ref{fig:delta_sigma1_mu} illustrates that $\Delta \log \Sigma_{1}$ correlates well with the other bulge type indicators, $\Delta \langle \mu_{e} \rangle$ and $\Delta \log \langle \Sigma_{e} \rangle$, with the latter showing a somewhat tighter correlation than the former. In both cases, the scatter is larger at the low-density end. Nevertheless, $\Delta \log \Sigma_{1}$ yields greater than 80\% consistency in bulge classification compared to the other two criteria.

  We examine how $\Delta\log\Sigma_{1}$ is distributed in the bulge color--magnitude diagram. The average separation line between classical and pseudo bulges classified using $\Delta \log \Sigma_{1}$ (Figure~\ref{fig:col_mag_mass}) is more or less consistent with that delineated using the mass-based Kormendy relation, $\Delta\log\langle\Sigma_e\rangle$. Although the average trend is similar, $\Delta\log\Sigma_1$ leads to greater tension between the structural and stellar population properties of bulges. This is evident when comparing the left panels of Figure~\ref{fig:col_mag_mass} in the faint end of bulges: there are more bulges with positive $\Delta\log\Sigma_1$ in the lower panel than those with positive $\Delta\log\langle\Sigma_e\rangle$ in the upper panel, both of which are outliers to the systematic trend. Namely, there would be more blue and faint classical bulges if the classifications were based on $\Sigma_1$. This is to be expected. After all, $\Sigma_{1}$ is a galaxy-wide parameter, whereas $\langle\Sigma_{e}\rangle$ is tailored specifically for the bulge component.

  \section{Discussion}
  \label{sec:discussion}

  \subsection{Effective Criteria for Bulge Classification}
  \label{sec:robustn-bulge-class}

  \citet{2017A&A+Neumann} found that classifications made using only the Kormendy relation are consistent with those obtained using multiple single-band photometric or stellar kinematic criteria. In our previous work, this was the major motivation for promoting the usage of the Kormendy relation to classify bulges \citep{2020ApJS+Gao}.  The more accurate bulge-to-disk decomposition afforded by the detailed two-dimensional analysis of CGS images revealed that the distribution of bulge \sersic{} indices is, in fact, not bimodal, casting doubt on the common practice of classifying bulges by the boundary of $n = 2$ (e.g., \citealt{2008AJ+Fisher}).  Also, classical and pseudo bulges overlap significantly in $B/T$ (\citealp{2020ApJS+Gao}; cf.\ Figure~\ref{fig:hist_comp}). In Section~\ref{sec:urge-replace-surface}, we showed that in the color--magnitude diagram of bulges $\Delta\langle\mu_e\rangle$ on average effectively sorts the bulge population into a smooth, monotonic sequence from red, luminous systems to blue, faint systems (Figure~\ref{fig:col_mag}). The performance of $B/T$ is inferior, and $n$ fares the worst. This reinforces our previous conclusion that the Kormendy relation is more effective than any single photometric structural parameter in classifying bulges.  Converting the Kormendy relation to a parameter space based on stellar mass offered us an additional insight: the bulk of the bulge classifications remained unchanged, and, so too, the overall statistics of the two bulge types.  We conclude that bulge classifications based on the Kormendy relation are robust against stellar population effects.

  Still, working with stellar mass brings other advantages that should not be overlooked. Classifying bulges using the stellar mass-based Kormendy relation leads to better agreement between the structural properties and stellar populations of classical and pseudo bulges. This is evident by inspecting Figures~\ref{fig:col_mag} and \ref{fig:col_mag_mass}: the mass-based parameters recognize more optically red bulges as classical bulges. This is, of course, expected. The selection criteria aim to pick denser bulges as classical bulges, and red bulges have higher $M/L$. Moreover, our mass-based method evidently identifies a much larger population of ultra-dense bulges (19 vs.\ 1; Section~\ref{sec:ultra-dense-bulges}), intriguing objects whose origin will be discussed more extensively in Section~\ref{sec:origin-ultra-dense}.  Lastly, the $\Sigma_1$ parameter is a cheap but effective parameter for bulge studies. We confirm previous studies that using $\Sigma_1$ to classify bulges not only yields good consistency with both the original and the mass-based Kormendy relation, but that $\Sigma_1$ also correlates with the two bulge indicators ($\Delta\langle\mu_e\rangle$ and $\Delta\langle\Sigma_e\rangle$) very well.  In other words, they can be used to predict each other. A tremendous asset is that measuring $\Sigma_1$ does not require bulge-to-disk decomposition whatsoever, and it also places less stringent demand on image resolution.

  Since our sample is not complete at low masses, we lack sufficient dynamic range to fully explore the bimodality of bulges (e.g., Figure~\ref{fig:mass_sigma1}) and the benefits of working with stellar mass. Given the prevalence of multiband images, however, we recommend working with the more physically meaningful mass parameters to study bulges whenever possible.

  \subsection{Ultra-dense Bulges}
  \label{sec:origin-ultra-dense}

\begin{figure*}
    \epsscale{1.18}
    \plotone{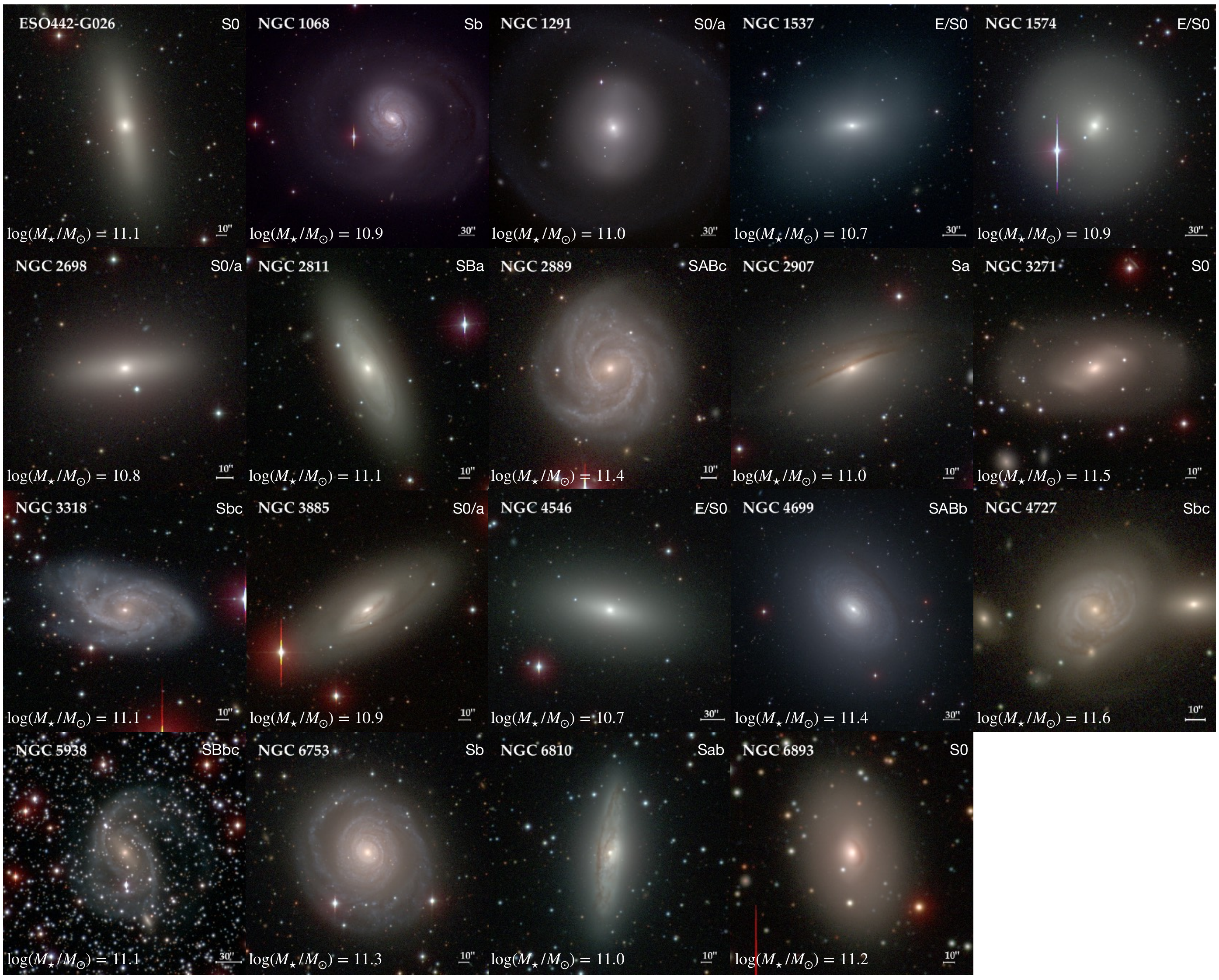}
    \caption{Atlas of the galaxies that host ultra-dense bulges. Their stellar masses and Hubble types from HyperLeda \citep{2003A&A+Paturel} are labeled in the lower-left and upper-right corner of each panel, respectively. \label{fig:UDB_atlas}}
  \end{figure*}

  In Section~\ref{sec:ultra-dense-bulges}, we designated a subsample of 19 classical bulges as ultra-dense bulges, whose $\langle\Sigma_e\rangle$ are upper $3\sigma$ outliers of the best-fit mass-based Kormendy relation of the CGS ellipticals. Here, we summarize their properties.  While they have the highest $\Sigma_1$ of all galaxies (Figure~\ref{fig:mass_sigma1}), as expected from their definition, ultra-dense bulges are not necessarily the most concentrated or most luminous members among the classical bulges (Figure~\ref{fig:col_mag}).  The similarity of their $\langle\Sigma_e\rangle$ with high-$z$ ($\ga 1.5$) ETGs, coupled with their very red optical colors at fixed bulge luminosity, hints of an early formation epoch. With a median total stellar mass of $10^{11.1}\,M_{\sun}$, their hosts are the most massive among all disk galaxies, as massive as the ellipticals in CGS (median stellar mass of $10^{11.2}\,M_{\sun}$). Two-thirds of them have early Hubble types (S0--Sa).  The $BVI$ three-color CGS images of the host galaxies are displayed in Figure~\ref{fig:UDB_atlas}.

  The above-mentioned characteristics of ultra-dense bulges are reminiscent of previously discovered relic high-$z$ massive compact galaxies \citep{2014ApJ+Trujillo, 2017MNRAS+Ferre-Mateu,2017MNRAS+Yildirim}. These high-$z$ survivors, fortunate enough to have escaped the violence of hierarchical merging, remain as dense and compact as high-$z$ ETGs at $z\approx 0$, while most of the other massive compact galaxies have experienced significant size growth through dry, minor (but see \citealt{2017ApJ+Davari}) mergers to become present-day massive ellipticals \citep[e.g.,][]{2009ApJ+Bezanson, 2013ApJ+Huang2, 2016ApJ+Huang, 2021ApJ+Zhu}. The ultra-dense bulges highlighted here, with median $\log\left(M_{\star}/M_{\sun}\right) = 10.3$ and median $R_e = 0.4\,\mathrm{kpc}$, bear a close resemblance to the population of similarly massive, dense bulges found lurking in the centers of nearby disk galaxies \citep{2015ApJ+Graham, 2016MNRAS+de_la_Rosa, 2020ApJ+Costantin}.  \citet{2021ApJ+Costantin, 2022ApJ+Costantin} argued that this subset of bulges descended directly from highly dissipative structures associated with an early population of galaxies formed at $z\gtrsim 3$.

  By analogy, can we hypothesize that the ultra-dense bulges are relic progenitors of the other classical bulges? It is not likely the case, because their hosts are more massive and have higher $\Sigma_1$ than many of the other hosts of classical bulges. Despite some effects that could reduce it, $\Sigma_1$ generally increases as galaxies evolve. Assuming that the ultra-dense bulges are relic high-$z$ ETGs, $\Sigma_1$ of their hosts should barely increase, yet their $\Sigma_1$ are still higher at $z\approx 0$ than the other classical bulge hosts. Thus, it is safe to conclude that the progenitors of the ultra-dense bulge hosts and the majority of the other classical bulge hosts are different in terms of stellar masses and/or central densities, at least already at the formation epoch of the ultra-dense bulges. Thus, the CGS ultra-dense bulges are probably relic high-$z$ ETGs, but their characteristics are not representative of the progenitors of most present-day classical bulges. A sample that is more complete in the lower stellar mass regime may help to clarify what the progenitors of other classical bulges look like.

  Having tried to link the ultra-dense bulges to structures that were assembled quickly at early times, it is also worthwhile to mention other possibilities. We note that six of the galaxies that host ultra-dense bulges were recognized to have buckled bars by \citeauthor{2017ApJ+Li}~(\citeyear{2017ApJ+Li}; NGC~1291,~2811,~3271,~3885,~4699,~and~4727). In addition, NGC~6893 seems to host a barlens, although it not was recognized by \citet{2017ApJ+Li}. The stellar mass range of the ultra-dense bulge hosts ($\ga 10^{11}\,M_\sun$) corresponds to the range where the fraction of buckled bars reaches its maximum ($\sim 40\%$) in the CGS disk galaxy sample \citep{2017ApJ+Li}. When a bar gets buckled, its stellar mass in the thickened part will be modeled as part of the photometric bulge in our decomposition strategy, thereby potentially increasing the measured average mass surface density within the half-light radius of the bulge. Moreover, the resultant bulge is expected to have an old stellar population, because the stellar population of bars in early-type disks is often old \citep{1996ASPC+Phillips}, the buckling process does not involve new star formation \citep{1990A&A+Combes}, and buckled bars could subsequently reduce gas inflow to the galaxy center \citep{2016MNRAS+Fragkoudi}. In light of these factors, the bar buckling may be a viable mechanism for forming the red ultra-dense bulges found in this study. We also find that some galaxies (NGC~1068,~1291,~3271,~4699,~and~6753; \citealt{2019ApJS+Gao}) host nuclear bars and/or rings. The presence of these features implies the presence of an underlying kinematically cold structure, namely a nuclear disk, which is a product of secular processes. However, \citet{2019MNRAS+de_Lorenzo-Caceres1} claimed, on the basis of structural properties and colors of their bulges, that most galaxies that have both a nuclear and a main bar host classical bulges. Moreover, the stellar population of nuclear disks is not necessarily young (e.g., NGC~1291; \citealt{2019MNRAS+de_Lorenzo-Caceres2, 2021A&A+Bittner}).

  To summarize: both early, rapid collapse and protracted secular evolution can contribute to the formation of ultra-dense bulges. If secular processes do play a role, it is difficult to understand why their influence is not more widespread in other bulges. Additional data and analysis more refined than that attempted here will be required to disentangle the contribution of various mechanisms.


  \section{Summary}
  \label{sec:summary}

  We measured optical colors of bulges for a sample of 312 CGS disk galaxies and converted our previous measurements of bulge structural parameters from flux to stellar mass units. We constructed the mass-based Kormendy relation of bulges and ellipticals and applied it to classify bulge types. Comparison of the classifications based on the original Kormendy relation and the mass-based one showed that the classifications of bulge types are consistent (to better than $90\%$) and the overall statistics of classical and pseudo bulges are barely changed. We conclude that the Kormendy relation as a classification tool is robust against stellar population effects. However, working with mass allows for better agreement between bulge structural properties and stellar populations, as it allows more bulges with old stellar population to be classified as classical bulges. It also helps to reveal a more significant population of ultra-dense bulges, which are probably relic primordial bulges as dense as high-$z$ ETGs, although secular processes may also contribute to their growth. We confirmed previous studies that $\Sigma_1$ is well correlated with bulge surface densities.

  \acknowledgments
  This work was supported by the National Science Foundation of China (11721303, 11991052, 12011540375, 12122301) and the China Manned Space Project (CMS-CSST-2021-A04, CMS-CSST-2021-A06). ZYL is supported by a Shanghai Natural Science Research Grant (21ZR1430600), and by the ``111'' project of the Ministry of Education under grant No.~B20019. Kavli IPMU was established by World Premier International Research Center Initiative (WPI), MEXT, Japan. We thank Aaron~Barth for helpful suggestions and for his contributions to the Carnegie-Irvine Galaxy Survey. HG thanks Hassen~Yesuf for useful discussions. We thank the referee for helpful comments that improved the manuscript.

  \bibliographystyle{aasjournal}
  \bibliography{myref}

\begin{thebibliography}{}
\expandafter\ifx\csname natexlab\endcsname\relax\def\natexlab#1{#1}\fi
\providecommand{\url}[1]{\href{#1}{#1}}

\bibitem[{{Bamford} {et~al.}(2011){Bamford}, {H{\"a}u{\ss}ler}, {Rojas}, \&
  {Borch}}]{2011ASPC+Bamford}
{Bamford}, S.~P., {H{\"a}u{\ss}ler}, B., {Rojas}, A., \& {Borch}, A. 2011, in
  ASP Conf. Ser. 442, Astronomical Data Analysis Software and Systems XX, ed.
  I.~N. {Evans}, A.~{Accomazzi}, D.~J. {Mink}, \& A.~H. {Rots} (San Francisco,
  CA: ASP), 479

\bibitem[{{Bell}(2008)}]{2008ApJ+Bell}
{Bell}, E.~F. 2008, \apj, 682, 355

\bibitem[{{Bell} {et~al.}(2003){Bell}, {McIntosh}, {Katz}, \&
  {Weinberg}}]{2003ApJS+Bell}
{Bell}, E.~F., {McIntosh}, D.~H., {Katz}, N., \& {Weinberg}, M.~D. 2003, \apjs,
  149, 289

\bibitem[{{Bell} {et~al.}(2012){Bell}, {van der Wel}, {Papovich}, {Kocevski},
  {Lotz}, {McIntosh}, {Kartaltepe}, {Faber}, {Ferguson}, {Koekemoer}, {Grogin},
  {Wuyts}, {Cheung}, {Conselice}, {Dekel}, {Dunlop}, {Giavalisco},
  {Herrington}, {Koo}, {McGrath}, {de Mello}, {Rix}, {Robaina}, \&
  {Williams}}]{2012ApJ+Bell}
{Bell}, E.~F., {van der Wel}, A., {Papovich}, C., {et~al.} 2012, \apj, 753, 167

\bibitem[{{Bender} {et~al.}(1992){Bender}, {Burstein}, \&
  {Faber}}]{1992ApJ+Bender}
{Bender}, R., {Burstein}, D., \& {Faber}, S.~M. 1992, \apj, 399, 462

\bibitem[{{Bezanson} {et~al.}(2009){Bezanson}, {van Dokkum}, {Tal},
  {Marchesini}, {Kriek}, {Franx}, \& {Coppi}}]{2009ApJ+Bezanson}
{Bezanson}, R., {van Dokkum}, P.~G., {Tal}, T., {et~al.} 2009, \apj, 697, 1290

\bibitem[{{Binggeli}(1994)}]{1994ESOC+Binggeli}
{Binggeli}, B. 1994, in ESO/OHP Workshop on Dwarf Galaxies, ed. G.~{Meylan} \&
  P.~{Prugniel} (Garching: ESO), 13

\bibitem[{{Bittner} {et~al.}(2021){Bittner}, {de Lorenzo-C{\'a}ceres},
  {Gadotti}, {S{\'a}nchez-Bl{\'a}zquez}, {Neumann}, {Coelho},
  {Falc{\'o}n-Barroso}, {Fragkoudi}, {Kim}, {Mart{\'\i}n-Navarro},
  {M{\'e}ndez-Abreu}, {P{\'e}rez}, {Querejeta}, \& {van de
  Ven}}]{2021A&A+Bittner}
{Bittner}, A., {de Lorenzo-C{\'a}ceres}, A., {Gadotti}, D.~A., {et~al.} 2021,
  \aap, 646, A42

\bibitem[{{Bournaud}(2016)}]{2016ASSL+Bournaud}
{Bournaud}, F. 2016, in Galactic Bulges, ed. E.~{Laurikainen}, R.~{Peletier},
  \& D.~{Gadotti} (New York: Springer), 355

\bibitem[{{Cappellari} {et~al.}(2012){Cappellari}, {McDermid}, {Alatalo},
  {Blitz}, {Bois}, {Bournaud}, {Bureau}, {Crocker}, {Davies}, {Davis}, {de
  Zeeuw}, {Duc}, {Emsellem}, {Khochfar}, {Krajnovi{\'c}}, {Kuntschner},
  {Lablanche}, {Morganti}, {Naab}, {Oosterloo}, {Sarzi}, {Scott}, {Serra},
  {Weijmans}, \& {Young}}]{2012Natur+Cappellari}
{Cappellari}, M., {McDermid}, R.~M., {Alatalo}, K., {et~al.} 2012, \nat, 484,
  485

\bibitem[{{Cappellari} {et~al.}(2013{\natexlab{a}}){Cappellari}, {Scott},
  {Alatalo}, {Blitz}, {Bois}, {Bournaud}, {Bureau}, {Crocker}, {Davies},
  {Davis}, {de Zeeuw}, {Duc}, {Emsellem}, {Khochfar}, {Krajnovi{\'c}},
  {Kuntschner}, {McDermid}, {Morganti}, {Naab}, {Oosterloo}, {Sarzi}, {Serra},
  {Weijmans}, \& {Young}}]{2013MNRAS+Cappellari1}
{Cappellari}, M., {Scott}, N., {Alatalo}, K., {et~al.} 2013{\natexlab{a}},
  \mnras, 432, 1709

\bibitem[{{Cappellari} {et~al.}(2013{\natexlab{b}}){Cappellari}, {McDermid},
  {Alatalo}, {Blitz}, {Bois}, {Bournaud}, {Bureau}, {Crocker}, {Davies},
  {Davis}, {de Zeeuw}, {Duc}, {Emsellem}, {Khochfar}, {Krajnovi{\'c}},
  {Kuntschner}, {Morganti}, {Naab}, {Oosterloo}, {Sarzi}, {Scott}, {Serra},
  {Weijmans}, \& {Young}}]{2013MNRAS+Cappellari2}
{Cappellari}, M., {McDermid}, R.~M., {Alatalo}, K., {et~al.}
  2013{\natexlab{b}}, \mnras, 432, 1862

\bibitem[{{Cheung} {et~al.}(2012){Cheung}, {Faber}, {Koo}, {Dutton}, {Simard},
  {McGrath}, {Huang}, {Bell}, {Dekel}, {Fang}, {Salim}, {Barro}, {Bundy},
  {Coil}, {Cooper}, {Conselice}, {Davis}, {Dom{\'\i}nguez}, {Kassin},
  {Kocevski}, {Koekemoer}, {Lin}, {Lotz}, {Newman}, {Phillips}, {Rosario},
  {Weiner}, \& {Willmer}}]{2012ApJ+Cheung}
{Cheung}, E., {Faber}, S.~M., {Koo}, D.~C., {et~al.} 2012, \apj, 760, 131

\bibitem[{Cleveland \& Devlin(1988)}]{1988J.AM.STAT.ASSOC+Cleveland}
Cleveland, W.~S., \& Devlin, S.~J. 1988, Journal of the American Statistical
  Association, 83, 596

\bibitem[{{Combes} {et~al.}(1990){Combes}, {Debbasch}, {Friedli}, \&
  {Pfenniger}}]{1990A&A+Combes}
{Combes}, F., {Debbasch}, F., {Friedli}, D., \& {Pfenniger}, D. 1990, \aap,
  233, 82

\bibitem[{{Combes} \& {Sanders}(1981)}]{1981A&A+Combes}
{Combes}, F., \& {Sanders}, R.~H. 1981, \aap, 96, 164

\bibitem[{{Conselice} {et~al.}(2011){Conselice}, {Bluck}, {Buitrago}, {Bauer},
  {Gr{\"u}tzbauch}, {Bouwens}, {Bevan}, {Mortlock}, {Dickinson}, {Daddi},
  {Yan}, {Scott}, {Chapman}, {Chary}, {Ferguson}, {Giavalisco}, {Grogin},
  {Illingworth}, {Jogee}, {Koekemoer}, {Lucas}, {Mobasher}, {Moustakas},
  {Papovich}, {Ravindranath}, {Siana}, {Teplitz}, {Trujillo}, {Urry}, \&
  {Weinzirl}}]{2011MNRAS+Conselice}
{Conselice}, C.~J., {Bluck}, A.~F.~L., {Buitrago}, F., {et~al.} 2011, \mnras,
  413, 80

\bibitem[{{Costantin} {et~al.}(2020){Costantin}, {M{\'e}ndez-Abreu}, {Corsini},
  {Morelli}, {de Lorenzo-C{\'a}ceres}, {Pagotto}, {Cuomo}, {Aguerri}, \&
  {Rubino}}]{2020ApJ+Costantin}
{Costantin}, L., {M{\'e}ndez-Abreu}, J., {Corsini}, E.~M., {et~al.} 2020,
  \apjl, 889, L3

\bibitem[{{Costantin} {et~al.}(2021){Costantin}, {P{\'e}rez-Gonz{\'a}lez},
  {M{\'e}ndez-Abreu}, {Huertas-Company}, {Dimauro}, {Alcalde-Pampliega},
  {Buitrago}, {Ceverino}, {Daddi}, {Dom{\'\i}nguez-S{\'a}nchez},
  {Espino-Briones}, {Hern{\'a}n-Caballero}, {Koekemoer}, \&
  {Rodighiero}}]{2021ApJ+Costantin}
{Costantin}, L., {P{\'e}rez-Gonz{\'a}lez}, P.~G., {M{\'e}ndez-Abreu}, J.,
  {et~al.} 2021, \apj, 913, 125

\bibitem[{{Costantin} {et~al.}(2022){Costantin}, {P{\'e}rez-Gonz{\'a}lez},
  {M{\'e}ndez-Abreu}, {Huertas-Company}, {Pampliega}, {Balcells}, {Barro},
  {Ceverino}, {Dimauro}, {S{\'a}nchez}, {Espino-Briones}, \&
  {Koekemoer}}]{2022ApJ+Costantin}
{Costantin}, L., {P{\'e}rez-Gonz{\'a}lez}, P.~G., {M{\'e}ndez-Abreu}, J.,
  {et~al.} 2022, \apj, 929, 121

\bibitem[{{Damjanov} {et~al.}(2011){Damjanov}, {Abraham}, {Glazebrook},
  {McCarthy}, {Caris}, {Carlberg}, {Chen}, {Crampton}, {Green}, {J{\o}rgensen},
  {Juneau}, {Le Borgne}, {Marzke}, {Mentuch}, {Murowinski}, {Roth}, {Savaglio},
  \& {Yan}}]{2011ApJ+Damjanov}
{Damjanov}, I., {Abraham}, R.~G., {Glazebrook}, K., {et~al.} 2011, \apjl, 739,
  L44

\bibitem[{{Davari} {et~al.}(2017){Davari}, {Ho}, {Mobasher}, \&
  {Canalizo}}]{2017ApJ+Davari}
{Davari}, R.~H., {Ho}, L.~C., {Mobasher}, B., \& {Canalizo}, G. 2017, \apj,
  836, 75

\bibitem[{{de la Rosa} {et~al.}(2016){de la Rosa}, {La Barbera}, {Ferreras},
  {S{\'a}nchez Almeida}, {Dalla Vecchia}, {Mart{\'\i}nez-Valpuesta}, \&
  {Stringer}}]{2016MNRAS+de_la_Rosa}
{de la Rosa}, I.~G., {La Barbera}, F., {Ferreras}, I., {et~al.} 2016, \mnras,
  457, 1916

\bibitem[{{de Lorenzo-C{\'a}ceres} {et~al.}(2019{\natexlab{a}}){de
  Lorenzo-C{\'a}ceres}, {M{\'e}ndez-Abreu}, {Thorne}, \&
  {Costantin}}]{2019MNRAS+de_Lorenzo-Caceres1}
{de Lorenzo-C{\'a}ceres}, A., {M{\'e}ndez-Abreu}, J., {Thorne}, B., \&
  {Costantin}, L. 2019{\natexlab{a}}, \mnras, 484, 665

\bibitem[{{de Lorenzo-C{\'a}ceres} {et~al.}(2019{\natexlab{b}}){de
  Lorenzo-C{\'a}ceres}, {S{\'a}nchez-Bl{\'a}zquez}, {M{\'e}ndez-Abreu},
  {Gadotti}, {Falc{\'o}n-Barroso}, {Mart{\'\i}nez-Valpuesta}, {Coelho},
  {Fragkoudi}, {Husemann}, {Leaman}, {P{\'e}rez}, {Querejeta}, {Seidel}, \&
  {van de Ven}}]{2019MNRAS+de_Lorenzo-Caceres2}
{de Lorenzo-C{\'a}ceres}, A., {S{\'a}nchez-Bl{\'a}zquez}, P.,
  {M{\'e}ndez-Abreu}, J., {et~al.} 2019{\natexlab{b}}, \mnras, 484, 5296

\bibitem[{{Djorgovski} \& {Davis}(1987)}]{1987ApJ+Djorgovski}
{Djorgovski}, S., \& {Davis}, M. 1987, \apj, 313, 59

\bibitem[{{Djorgovski} {et~al.}(1988){Djorgovski}, {de Carvalho}, \&
  {Han}}]{1988ASPC+Djorgovski}
{Djorgovski}, S., {de Carvalho}, R., \& {Han}, M.-S. 1988, in ASP Conf. Ser. 4,
  The Extragalactic Distance Scale, ed. S.~{van den Bergh} \& C.~J. {Pritchet}
  (San Francisco, CA: ASP), 329

\bibitem[{{D'Onofrio} {et~al.}(2017){D'Onofrio}, {Cariddi}, {Chiosi}, {Chiosi},
  \& {Marziani}}]{2017ApJ+DOnofrio}
{D'Onofrio}, M., {Cariddi}, S., {Chiosi}, C., {Chiosi}, E., \& {Marziani}, P.
  2017, \apj, 838, 163

\bibitem[{{D'Onofrio} {et~al.}(2013){D'Onofrio}, {Fasano}, {Moretti},
  {Marziani}, {Bindoni}, {Fritz}, {Varela}, {Bettoni}, {Cava}, {Poggianti},
  {Gullieuszik}, {Kj{\ae}rgaard}, {Moles}, {Vulcani}, {Omizzolo}, {Couch}, \&
  {Dressler}}]{2013MNRAS+DOnofrio}
{D'Onofrio}, M., {Fasano}, G., {Moretti}, A., {et~al.} 2013, \mnras, 435, 45

\bibitem[{{Dressler} {et~al.}(1987){Dressler}, {Lynden-Bell}, {Burstein},
  {Davies}, {Faber}, {Terlevich}, \& {Wegner}}]{1987ApJ+Dressler}
{Dressler}, A., {Lynden-Bell}, D., {Burstein}, D., {et~al.} 1987, \apj, 313, 42

\bibitem[{{Faber} {et~al.}(1987){Faber}, {Dressler}, {Davies}, {Burstein}, \&
  {Lynden-Bell}}]{1987nngp+Faber}
{Faber}, S.~M., {Dressler}, A., {Davies}, R.~L., {Burstein}, D., \&
  {Lynden-Bell}, D. 1987, in Nearly Normal Galaxies. From the Planck Time to
  the Present, ed. S.~M. {Faber} (New York: Springer), 175

\bibitem[{{Faber} \& {Jackson}(1976)}]{1976ApJ+Faber2}
{Faber}, S.~M., \& {Jackson}, R.~E. 1976, \apj, 204, 668

\bibitem[{{Faber} {et~al.}(1997){Faber}, {Tremaine}, {Ajhar}, {Byun},
  {Dressler}, {Gebhardt}, {Grillmair}, {Kormendy}, {Lauer}, \&
  {Richstone}}]{1997AJ+Faber}
{Faber}, S.~M., {Tremaine}, S., {Ajhar}, E.~A., {et~al.} 1997, \aj, 114, 1771

\bibitem[{{Fang} {et~al.}(2013){Fang}, {Faber}, {Koo}, \&
  {Dekel}}]{2013ApJ+Fang}
{Fang}, J.~J., {Faber}, S.~M., {Koo}, D.~C., \& {Dekel}, A. 2013, \apj, 776, 63

\bibitem[{{Ferr{\'e}-Mateu} {et~al.}(2017){Ferr{\'e}-Mateu}, {Trujillo},
  {Mart{\'{\i}}n-Navarro}, {Vazdekis}, {Mezcua}, {Balcells}, \&
  {Dom{\'{\i}}nguez}}]{2017MNRAS+Ferre-Mateu}
{Ferr{\'e}-Mateu}, A., {Trujillo}, I., {Mart{\'{\i}}n-Navarro}, I., {et~al.}
  2017, \mnras, 467, 1929

\bibitem[{{Fisher} \& {Drory}(2008)}]{2008AJ+Fisher}
{Fisher}, D.~B., \& {Drory}, N. 2008, \aj, 136, 773

\bibitem[{{Fisher} \& {Drory}(2016)}]{2016ASSL+Fisher}
{Fisher}, D.~B., \& {Drory}, N. 2016, in Galactic Bulges, ed. E.~{Laurikainen},
  R.~{Peletier}, \& D.~{Gadotti} (New York: Springer), 41

\bibitem[{{Fisher} {et~al.}(2009){Fisher}, {Drory}, \&
  {Fabricius}}]{2009ApJ+Fisher}
{Fisher}, D.~B., {Drory}, N., \& {Fabricius}, M.~H. 2009, \apj, 697, 630

\bibitem[{{Fragkoudi} {et~al.}(2016){Fragkoudi}, {Athanassoula}, \&
  {Bosma}}]{2016MNRAS+Fragkoudi}
{Fragkoudi}, F., {Athanassoula}, E., \& {Bosma}, A. 2016, \mnras, 462, L41

\bibitem[{{Franx} {et~al.}(2008){Franx}, {van Dokkum}, {F{\"o}rster Schreiber},
  {Wuyts}, {Labb{\'e}}, \& {Toft}}]{2008ApJ+Franx}
{Franx}, M., {van Dokkum}, P.~G., {F{\"o}rster Schreiber}, N.~M., {et~al.}
  2008, \apj, 688, 770

\bibitem[{{Gadotti}(2009)}]{2009MNRAS+Gadotti}
{Gadotti}, D.~A. 2009, \mnras, 393, 1531

\bibitem[{{Gao} \& {Ho}(2017)}]{2017ApJ+Gao}
{Gao}, H., \& {Ho}, L.~C. 2017, \apj, 845, 114

\bibitem[{{Gao} {et~al.}(2018){Gao}, {Ho}, {Barth}, \& {Li}}]{2018ApJ+Gao}
{Gao}, H., {Ho}, L.~C., {Barth}, A.~J., \& {Li}, Z.-Y. 2018, \apj, 862, 100

\bibitem[{{Gao} {et~al.}(2019){Gao}, {Ho}, {Barth}, \& {Li}}]{2019ApJS+Gao}
{Gao}, H., {Ho}, L.~C., {Barth}, A.~J., \& {Li}, Z.-Y. 2019, \apjs, 244, 34

\bibitem[{{Gao} {et~al.}(2020){Gao}, {Ho}, {Barth}, \& {Li}}]{2020ApJS+Gao}
{Gao}, H., {Ho}, L.~C., {Barth}, A.~J., \& {Li}, Z.-Y. 2020, \apjs, 247, 20

\bibitem[{{Graham} {et~al.}(2015){Graham}, {Dullo}, \&
  {Savorgnan}}]{2015ApJ+Graham}
{Graham}, A.~W., {Dullo}, B.~T., \& {Savorgnan}, G.~A.~D. 2015, \apj, 804, 32

\bibitem[{{Grogin} {et~al.}(2011){Grogin}, {Kocevski}, {Faber}, {Ferguson},
  {Koekemoer}, {Riess}, {Acquaviva}, {Alexander}, {Almaini}, {Ashby}, {Barden},
  {Bell}, {Bournaud}, {Brown}, {Caputi}, {Casertano}, {Cassata}, {Castellano},
  {Challis}, {Chary}, {Cheung}, {Cirasuolo}, {Conselice}, {Roshan Cooray},
  {Croton}, {Daddi}, {Dahlen}, {Dav{\'e}}, {de Mello}, {Dekel}, {Dickinson},
  {Dolch}, {Donley}, {Dunlop}, {Dutton}, {Elbaz}, {Fazio}, {Filippenko},
  {Finkelstein}, {Fontana}, {Gardner}, {Garnavich}, {Gawiser}, {Giavalisco},
  {Grazian}, {Guo}, {Hathi}, {H{\"a}ussler}, {Hopkins}, {Huang}, {Huang},
  {Jha}, {Kartaltepe}, {Kirshner}, {Koo}, {Lai}, {Lee}, {Li}, {Lotz}, {Lucas},
  {Madau}, {McCarthy}, {McGrath}, {McIntosh}, {McLure}, {Mobasher},
  {Moustakas}, {Mozena}, {Nandra}, {Newman}, {Niemi}, {Noeske}, {Papovich},
  {Pentericci}, {Pope}, {Primack}, {Rajan}, {Ravindranath}, {Reddy}, {Renzini},
  {Rix}, {Robaina}, {Rodney}, {Rosario}, {Rosati}, {Salimbeni}, {Scarlata},
  {Siana}, {Simard}, {Smidt}, {Somerville}, {Spinrad}, {Straughn}, {Strolger},
  {Telford}, {Teplitz}, {Trump}, {van der Wel}, {Villforth}, {Wechsler},
  {Weiner}, {Wiklind}, {Wild}, {Wilson}, {Wuyts}, {Yan}, \&
  {Yun}}]{2011ApJS+Grogin}
{Grogin}, N.~A., {Kocevski}, D.~D., {Faber}, S.~M., {et~al.} 2011, \apjs, 197,
  35

\bibitem[{{Ho} {et~al.}(2011){Ho}, {Li}, {Barth}, {Seigar}, \&
  {Peng}}]{2011ApJS+Ho}
{Ho}, L.~C., {Li}, Z.-Y., {Barth}, A.~J., {Seigar}, M.~S., \& {Peng}, C.~Y.
  2011, \apjs, 197, 21

\bibitem[{{Huang} {et~al.}(2013){Huang}, {Ho}, {Peng}, {Li}, \&
  {Barth}}]{2013ApJ+Huang2}
{Huang}, S., {Ho}, L.~C., {Peng}, C.~Y., {Li}, Z.-Y., \& {Barth}, A.~J. 2013,
  \apjl, 768, L28

\bibitem[{{Huang} {et~al.}(2016){Huang}, {Ho}, {Peng}, {Li}, \&
  {Barth}}]{2016ApJ+Huang}
{Huang}, S., {Ho}, L.~C., {Peng}, C.~Y., {Li}, Z.-Y., \& {Barth}, A.~J. 2016,
  \apj, 821, 114

\bibitem[{{Hyde} \& {Bernardi}(2009)}]{2009MNRAS+Hyde}
{Hyde}, J.~B., \& {Bernardi}, M. 2009, \mnras, 396, 1171

\bibitem[{{Kauffmann} {et~al.}(2006){Kauffmann}, {Heckman}, {De Lucia},
  {Brinchmann}, {Charlot}, {Tremonti}, {White}, \&
  {Brinkmann}}]{2006MNRAS+Kauffmann}
{Kauffmann}, G., {Heckman}, T.~M., {De Lucia}, G., {et~al.} 2006, \mnras, 367,
  1394

\bibitem[{{Kauffmann} {et~al.}(2003){Kauffmann}, {Heckman}, {White}, {Charlot},
  {Tremonti}, {Peng}, {Seibert}, {Brinkmann}, {Nichol}, {SubbaRao}, \&
  {York}}]{2003MNRAS+Kauffmann}
{Kauffmann}, G., {Heckman}, T.~M., {White}, S. D.~M., {et~al.} 2003, \mnras,
  341, 54

\bibitem[{{Kennedy} {et~al.}(2016){Kennedy}, {Bamford}, {H{\"a}u{\ss}ler},
  {Baldry}, {Bremer}, {Brough}, {Brown}, {Driver}, {Duncan}, {Graham},
  {Holwerda}, {Hopkins}, {Kelvin}, {Lange}, {Phillipps}, {Vika}, \&
  {Vulcani}}]{2016MNRAS+Kennedy}
{Kennedy}, R., {Bamford}, S.~P., {H{\"a}u{\ss}ler}, B., {et~al.} 2016, \mnras,
  460, 3458

\bibitem[{Kormendy(1977)}]{1977ApJ+Kormendy2}
Kormendy, J. 1977, \apj, 218, 333

\bibitem[{{Kormendy}(1981)}]{1981seng.proc+Kormendy}
{Kormendy}, J. 1981, in The Structure and Evolution of Normal Galaxies, ed.
  S.~M. {Fall} \& D.~{Lynden-Bell} (Cambridge: Cambridge Univ. Press), 85

\bibitem[{{Kormendy}(1982)}]{1982SAAS+Kormendy}
{Kormendy}, J. 1982, in Twelfth Advanced Course of the Swiss Society of
  Astronomy and Astrophysics, Morphology and Dynamics of Galaxies, ed.
  L.~{Martinet} \& M.~{Mayor} (Sauverny: Geneva Obs.), 113

\bibitem[{{Kormendy}(1985)}]{1985ApJ+Kormendy}
{Kormendy}, J. 1985, \apj, 295, 73

\bibitem[{{Kormendy}(1987)}]{1987nngp+Kormendy}
{Kormendy}, J. 1987, in Nearly Normal Galaxies. From the Planck Time to the
  Present, ed. S.~M. {Faber} (New York: Springer), 163

\bibitem[{{Kormendy}(1993)}]{1993IAUS+Kormendy}
{Kormendy}, J. 1993, in IAU Symp. 153, Galactic Bulges, ed. H.~{Dejonghe} \&
  H.~J. {Habing} (Dordrecht: Kluwer), 209

\bibitem[{{Kormendy} {et~al.}(2010){Kormendy}, {Drory}, {Bender}, \&
  {Cornell}}]{2010ApJ+Kormendy2}
{Kormendy}, J., {Drory}, N., {Bender}, R., \& {Cornell}, M.~E. 2010, \apj, 723,
  54

\bibitem[{{Kormendy} \& {Fisher}(2008)}]{2008ASPC+Kormendy}
{Kormendy}, J., \& {Fisher}, D.~B. 2008, in ASP Conf. Ser. 396, Formation and
  Evolution of Galaxy Disks, ed. J.~G. {Funes} \& E.~M. {Corsini} (San
  Francisco, CA: ASP), 297

\bibitem[{{Kormendy} \& {Kennicutt}(2004)}]{2004ARA&A+Kormendy}
{Kormendy}, J., \& {Kennicutt}, Jr., R.~C. 2004, \araa, 42, 603

\bibitem[{{Kroupa}(2001)}]{2001MNRAS+Kroupa}
{Kroupa}, P. 2001, \mnras, 322, 231

\bibitem[{{Li} {et~al.}(2017){Li}, {Ho}, \& {Barth}}]{2017ApJ+Li}
{Li}, Z.-Y., {Ho}, L.~C., \& {Barth}, A.~J. 2017, \apj, 845, 87

\bibitem[{{Li} {et~al.}(2011){Li}, {Ho}, {Barth}, \& {Peng}}]{2011ApJS+Li}
{Li}, Z.-Y., {Ho}, L.~C., {Barth}, A.~J., \& {Peng}, C.~Y. 2011, \apjs, 197, 22

\bibitem[{{Luo} {et~al.}(2020){Luo}, {Faber}, {Rodr{\'\i}guez-Puebla}, {Woo},
  {Guo}, {Koo}, {Primack}, {Chen}, {Yesuf}, {Lin}, {Barro}, {Fang}, {Pand ya},
  {Huertas-Company}, \& {Mao}}]{2020MNRAS+Luo}
{Luo}, Y., {Faber}, S.~M., {Rodr{\'\i}guez-Puebla}, A., {et~al.} 2020, \mnras,
  493, 1686

\bibitem[{{Lynden-Bell} {et~al.}(1988){Lynden-Bell}, {Faber}, {Burstein},
  {Davies}, {Dressler}, {Terlevich}, \& {Wegner}}]{1988ApJ+Lynden-Bell}
{Lynden-Bell}, D., {Faber}, S.~M., {Burstein}, D., {et~al.} 1988, \apj, 326, 19

\bibitem[{{Neumann} {et~al.}(2017){Neumann}, {Wisotzki}, {Choudhury},
  {Gadotti}, {Walcher}, {Bland-Hawthorn}, {Garc{\'{\i}}a-Benito}, {Gonz{\'a}lez
  Delgado}, {Husemann}, {Marino}, {M{\'a}rquez}, {S{\'a}nchez}, {Ziegler}, \&
  {Califa Collaboration}}]{2017A&A+Neumann}
{Neumann}, J., {Wisotzki}, L., {Choudhury}, O.~S., {et~al.} 2017, \aap, 604,
  A30

\bibitem[{{Papovich} {et~al.}(2012){Papovich}, {Bassett}, {Lotz}, {van der
  Wel}, {Tran}, {Finkelstein}, {Bell}, {Conselice}, {Dekel}, {Dunlop}, {Guo},
  {Faber}, {Farrah}, {Ferguson}, {Finkelstein}, {H{\"a}ussler}, {Kocevski},
  {Koekemoer}, {Koo}, {McGrath}, {McLure}, {McIntosh}, {Momcheva}, {Newman},
  {Rudnick}, {Weiner}, {Willmer}, \& {Wuyts}}]{2012ApJ+Papovich}
{Papovich}, C., {Bassett}, R., {Lotz}, J.~M., {et~al.} 2012, \apj, 750, 93

\bibitem[{{Paturel} {et~al.}(2003){Paturel}, {Petit}, {Prugniel}, {Theureau},
  {Rousseau}, {Brouty}, {Dubois}, \& {Cambr{\'e}sy}}]{2003A&A+Paturel}
{Paturel}, G., {Petit}, C., {Prugniel}, P., {et~al.} 2003, \aap, 412, 45

\bibitem[{{Peebles} \& {Nusser}(2010)}]{2010Natur+Peebles}
{Peebles}, P.~J.~E., \& {Nusser}, A. 2010, \nat, 465, 565

\bibitem[{{Pfenniger} \& {Norman}(1990)}]{1990ApJ+Pfenniger}
{Pfenniger}, D., \& {Norman}, C. 1990, \apj, 363, 391

\bibitem[{{Phillips}(1996)}]{1996ASPC+Phillips}
{Phillips}, A.~C. 1996, in ASP Conf. Ser. 91, IAU Colloq. 157: Barred Galaxies,
  ed. R.~{Buta}, D.~A. {Crocker}, \& B.~G. {Elmegreen} (San Francisco, CA:
  ASP), 44

\bibitem[{{Renzini} \& {Ciotti}(1993)}]{1993ApJ+Renzini}
{Renzini}, A., \& {Ciotti}, L. 1993, \apjl, 416, L49

\bibitem[{{S{\'a}nchez-Bl{\'a}zquez}(2016)}]{2016ASSL+Sanchez-Blazquez}
{S{\'a}nchez-Bl{\'a}zquez}, P. 2016, in Galactic Bulges, ed. E.~{Laurikainen},
  R.~{Peletier}, \& D.~{Gadotti} (New York: Springer), 127

\bibitem[{{Scarlata} {et~al.}(2007){Scarlata}, {Carollo}, {Lilly}, {Sargent},
  {Feldmann}, {Kampczyk}, {Porciani}, {Koekemoer}, {Scoville}, {Kneib},
  {Leauthaud}, {Massey}, {Rhodes}, {Tasca}, {Capak}, {Maier}, {McCracken},
  {Mobasher}, {Renzini}, {Taniguchi}, {Thompson}, {Sheth}, {Ajiki}, {Aussel},
  {Murayama}, {Sanders}, {Sasaki}, {Shioya}, \&
  {Takahashi}}]{2007ApJS+Scarlata}
{Scarlata}, C., {Carollo}, C.~M., {Lilly}, S., {et~al.} 2007, \apjs, 172, 406

\bibitem[{{Scoville} {et~al.}(2007){Scoville}, {Abraham}, {Aussel}, {Barnes},
  {Benson}, {Blain}, {Calzetti}, {Comastri}, {Capak}, {Carilli}, {Carlstrom},
  {Carollo}, {Colbert}, {Daddi}, {Ellis}, {Elvis}, {Ewald}, {Fall},
  {Franceschini}, {Giavalisco}, {Green}, {Griffiths}, {Guzzo}, {Hasinger},
  {Impey}, {Kneib}, {Koda}, {Koekemoer}, {Lefevre}, {Lilly}, {Liu},
  {McCracken}, {Massey}, {Mellier}, {Miyazaki}, {Mobasher}, {Mould}, {Norman},
  {Refregier}, {Renzini}, {Rhodes}, {Rich}, {Sanders}, {Schiminovich},
  {Schinnerer}, {Scodeggio}, {Sheth}, {Shopbell}, {Taniguchi}, {Tyson}, {Urry},
  {Van Waerbeke}, {Vettolani}, {White}, \& {Yan}}]{2007ApJS+Scoville}
{Scoville}, N., {Abraham}, R.~G., {Aussel}, H., {et~al.} 2007, \apjs, 172, 38

\bibitem[{{Sellwood}(2014)}]{2014RvMP+Sellwood}
{Sellwood}, J.~A. 2014, RvMP, 86, 1

\bibitem[{{Sellwood} \& {Wilkinson}(1993)}]{1993RPPh+Sellwood}
{Sellwood}, J.~A., \& {Wilkinson}, A. 1993, RPPh, 56, 173

\bibitem[{{S{\'e}rsic}(1968)}]{1968adga+Sersic}
{S{\'e}rsic}, J.~L. 1968, {Atlas de Galaxias Australes} (C{\'o}rdoba: Obs.
  Astron., Univ. Nac. C{\'o}rdoba)

\bibitem[{{Szomoru} {et~al.}(2012){Szomoru}, {Franx}, \& {van
  Dokkum}}]{2012ApJ+Szomoru}
{Szomoru}, D., {Franx}, M., \& {van Dokkum}, P.~G. 2012, \apj, 749, 121

\bibitem[{{Tacchella} {et~al.}(2016){Tacchella}, {Dekel}, {Carollo},
  {Ceverino}, {DeGraf}, {Lapiner}, {Mand elker}, \&
  {Primack}}]{2016MNRAS+Tacchella}
{Tacchella}, S., {Dekel}, A., {Carollo}, C.~M., {et~al.} 2016, \mnras, 458, 242

\bibitem[{{Tacchella} {et~al.}(2019){Tacchella}, {Diemer}, {Hernquist},
  {Genel}, {Marinacci}, {Nelson}, {Pillepich}, {Rodriguez-Gomez}, {Sales},
  {Springel}, \& {Vogelsberger}}]{2019MNRAS+Tacchella}
{Tacchella}, S., {Diemer}, B., {Hernquist}, L., {et~al.} 2019, \mnras, 487,
  5416

\bibitem[{{Tasca} {et~al.}(2009){Tasca}, {Kneib}, {Iovino}, {Le F{\`e}vre},
  {Kova{\v{c}}}, {Bolzonella}, {Lilly}, {Abraham}, {Cassata}, {Cucciati},
  {Guzzo}, {Tresse}, {Zamorani}, {Capak}, {Garilli}, {Scodeggio}, {Sheth},
  {Zucca}, {Carollo}, {Contini}, {Mainieri}, {Renzini}, {Bardelli},
  {Bongiorno}, {Caputi}, {Coppa}, {de La Torre}, {de Ravel}, {Franzetti},
  {Kampczyk}, {Knobel}, {Koekemoer}, {Lamareille}, {Le Borgne}, {Le Brun},
  {Maier}, {Mignoli}, {Pello}, {Peng}, {Perez Montero}, {Ricciardelli},
  {Silverman}, {Vergani}, {Tanaka}, {Abbas}, {Bottini}, {Cappi}, {Cimatti},
  {Ilbert}, {Leauthaud}, {Maccagni}, {Marinoni}, {McCracken}, {Memeo},
  {Meneux}, {Oesch}, {Porciani}, {Pozzetti}, {Scaramella}, \&
  {Scarlata}}]{2009A&A+Tasca}
{Tasca}, L.~A.~M., {Kneib}, J.~P., {Iovino}, A., {et~al.} 2009, \aap, 503, 379

\bibitem[{{Toomre}(1977)}]{1977egsp+Tmoore}
{Toomre}, A. 1977, in Evolution of Galaxies and Stellar Populations, ed. B.~M.
  {Tinsley} \& R.~B. {Larson} (New Haven, CT: Yale Univ. Obs.), 401

\bibitem[{{Trujillo} {et~al.}(2014){Trujillo}, {Ferr{\'e}-Mateu}, {Balcells},
  {Vazdekis}, \& {S{\'a}nchez-Bl{\'a}zquez}}]{2014ApJ+Trujillo}
{Trujillo}, I., {Ferr{\'e}-Mateu}, A., {Balcells}, M., {Vazdekis}, A., \&
  {S{\'a}nchez-Bl{\'a}zquez}, P. 2014, \apjl, 780, L20

\bibitem[{{Vika} {et~al.}(2014){Vika}, {Bamford}, {H{\"a}u{\ss}ler}, \&
  {Rojas}}]{2014MNRAS+Vika}
{Vika}, M., {Bamford}, S.~P., {H{\"a}u{\ss}ler}, B., \& {Rojas}, A.~L. 2014,
  \mnras, 444, 3603

\bibitem[{{Woo} {et~al.}(2017){Woo}, {Carollo}, {Faber}, {Dekel}, \&
  {Tacchella}}]{2017MNRAS+Woo}
{Woo}, J., {Carollo}, C.~M., {Faber}, S.~M., {Dekel}, A., \& {Tacchella}, S.
  2017, \mnras, 464, 1077

\bibitem[{{Woo} \& {Ellison}(2019)}]{2019MNRAS+Woo}
{Woo}, J., \& {Ellison}, S.~L. 2019, \mnras, 487, 1927

\bibitem[{{Wuyts} {et~al.}(2011){Wuyts}, {F{\"o}rster Schreiber}, {van der
  Wel}, {Magnelli}, {Guo}, {Genzel}, {Lutz}, {Aussel}, {Barro}, {Berta},
  {Cava}, {Graci{\'a}-Carpio}, {Hathi}, {Huang}, {Kocevski}, {Koekemoer},
  {Lee}, {Le Floc'h}, {McGrath}, {Nordon}, {Popesso}, {Pozzi}, {Riguccini},
  {Rodighiero}, {Saintonge}, \& {Tacconi}}]{2011ApJ+Wuyts}
{Wuyts}, S., {F{\"o}rster Schreiber}, N.~M., {van der Wel}, A., {et~al.} 2011,
  \apj, 742, 96

\bibitem[{{Yesuf} {et~al.}(2020){Yesuf}, {Faber}, {Koo}, {Woo}, {Primack}, \&
  {Luo}}]{2020ApJ+Yesuf}
{Yesuf}, H.~M., {Faber}, S.~M., {Koo}, D.~C., {et~al.} 2020, \apj, 889, 14

\bibitem[{{Y{\i}ld{\i}r{\i}m} {et~al.}(2017){Y{\i}ld{\i}r{\i}m}, {van den
  Bosch}, {van de Ven}, {Mart{\'\i}n-Navarro}, {Walsh}, {Husemann},
  {G{\"u}ltekin}, \& {Gebhardt}}]{2017MNRAS+Yildirim}
{Y{\i}ld{\i}r{\i}m}, A., {van den Bosch}, R. C.~E., {van de Ven}, G., {et~al.}
  2017, \mnras, 468, 4216

\bibitem[{{Zhu} {et~al.}(2021){Zhu}, {Ho}, \& {Gao}}]{2021ApJ+Zhu}
{Zhu}, P., {Ho}, L.~C., \& {Gao}, H. 2021, \apj, 907, 6

\bibitem[{{Zolotov} {et~al.}(2015){Zolotov}, {Dekel}, {Mandelker}, {Tweed},
  {Inoue}, {DeGraf}, {Ceverino}, {Primack}, {Barro}, \&
  {Faber}}]{2015MNRAS+Zolotov}
{Zolotov}, A., {Dekel}, A., {Mandelker}, N., {et~al.} 2015, \mnras, 450, 2327

\end{thebibliography}
\end{CJK*}
\end{document}